\documentclass[aps,prd,reprint,showpacs,
groupedaddress,
nobibnotes,nofootinbib]{revtex4-1}
\usepackage{amsmath,amssymb,amsthm}
\usepackage{graphicx}
\usepackage{mathrsfs}
\usepackage{multirow}
\newcommand{\mn}{{\mu\nu}}

\newcommand{\beq}{\begin{equation}}
\newcommand{\eeq}{\end{equation}}
\newcommand{\bw}{\begin{widetext}}
\newcommand{\ew}{\end{widetext}}

\begin{document}
\title{The Finsler Type of Space-time Realization of Deformed Very Special Relativity}

\author{Lei Zhang}

\author{Xun Xue}
\email[Corresponding author: ]{xxue@phy.ecnu.edu.cn}

\affiliation{Institute of Theoretical Physics, Department of Physics, East China Normal University, No.500, Dongchuan Road, Shanghai 200241, China}

\affiliation{Kavli Institute for Theoretical Physics China at the Chinese Academy of Sciences,Beijing, China}

\begin{abstract}
We investigate here all the possible invariant metric functions under the action of various kinds of semi-direct product Poincar\'e subgroups and their deformed partners. The investigation exhausts the possible theoretical frameworks for the spacetime realization of Cohen-Glashow's very special relativity and the deformation very special relativity approach by Gibbons-Gomis-Pope. Within Finsler-Minkowski type of spacetime, we find that the spacetime emerge a Finsler type of geometry in most cases both for undermed Poincar\'e subgroup and for deformed one. We give an explanation that the rotation operation should be kept even in a Lorentz violating theory from geometrical view of point. We also find that the admissible geometry for $DTE3b$, $TE(2)$, $ISO(3)$ and $ISO(2,1)$ actually consists of a family in which the metric function vary with a freedom of arbitrary function of the specified combination of variables. The only principle for choosing the correct geometry from the family can only be the dynamical behavior of physics in the spacetime.
\end{abstract}

\maketitle

\section{Introduction}

The local Lorentz symmetry and CPT invariance is one of the fundamentals of modern physics. The theoretical investigation and experimental examination of Lorentz symmetry have made considerable progress and attracted a lot of attentions since the mid of 1990s \cite{Mattingly:2005re}.

There are many attempts to investigate the possible Lorentz violation from theoretical aspect. Coleman and Glashow consider the case of spacetime translations along with exact rotational symmetry in the rest frame of the cosmic background radiation, but allow small departures from boost invariance in this frame. They developed a perturbative  framework to investigate the deviation from Lorentz invariance in which the departure of Lorentz invariance is parametrized in terms of a fixed timelike fourvector or spurion \cite{Coleman:1998ti}.  Colladay and Kostelecky \cite{Kosteleck:1998lv} proposed the model incorporating Lorentz and CPT violation extension of the standard model by introducing into the Lagrangian of more general spurion-mediated perturbations, which are sometimes referred to as expectation values of Lorentz tensors following spontaneous Lorentz breaking. It is also argued that large boosts naturally uncover the structure of spacetime at arbitrary small scales and it is unclear how this could be conciliated with the existence of a fundamental scale for the quantum gravitational phenomena, i.e. the Planck scale. The modification of special relativity with an additional fundamental length scale, the Planck scale, is known as doubly special relativity(DSR) \cite{Amelino-Camelia:2000mn}. The realization of DSR can be noncommutative spacetime or the non-linear realization of Poincar\'e group \cite{Amelino-Camelia:2010di}. The main feature of these realization of DSR involves the deformed dispersion relation which can also leads to Finsler type of spacetime geometry \cite{Girelli:2007pg}.

Because at low energy scales, parity $P$, charge conjugation $C$ and time reversal $T$ are individually good symmetries of nature while there is evidence of $CP$ violation for higher energies, one may consider the possible failure of Poincar\'e symmetry at such high energy scales. One theoretical possibility is that the spacetime symmetry of all the observed physical phenomena might be some proper subgroups of the Lorentz group along with the spacetime translations only if these kind of proper subgroups of Poincar\'e group incorporating with either of the discrete operations $P$, $T$ $CP$ or $CT$, can be enlarged to the full Poincar\'e group. The generic models based on these smaller subgroups are restricted by the principle of Very Special Relativity (VSR), proposed by Cohen and Glashow \cite{Cohen:2006vy}.  Cohen and Glashow argued that the local symmetry of physics might not need to be as large as Lorentz group but its proper subgroup, while the full symmetry restores to Poincar\'e group when discrete symmetry $P$, $T$ or $CP$ enters. The Lorentz violation is thus connected with CP violation. Since CP violating effects in nature are small, it is possible that Lorentz-violating effects may be similarly small. They identified these VSR subgroups up to isomorphism as T(2) (2-dimensional translations) with generators $T_{1}=K_{x}+J_{y}$ and $T_{2} = K_{y} - J_{x}$, where $\mathbf{J}$ and $\mathbf{K}$ are the generators of rotations and boosts respectively, E(2) (3-parameter Euclidean motion) with generators $T_{1},T_{2}$ and $J_z$, HOM(2) (3-parameter orientation preserving transformations) with generators $T_{1},T_{2}$ and $K_z$ and SIM(2) (4-parameter similitude group)with generators $T_{1},T_{2}, J_z$ and $K_z$. The spurion strategy can also be applied to VSP. The invariant tensor for group $E(2)$ can be a 4-vector $n=(1,0,0,1)$ while the symmetry groups $T(2)$ admits many invariant tensors. There is neither invariant tensors for $HOM(2)$ and $SIM(2)$ nor the local Lorentz symmetry breaking perturbative discription for either of these groups.

Concerning how to realize VSR, Sheikh-Jabbar et.al proved that the quantum field theory on the noncommutative Moyal plane with light-like noncommutativity possesses VSR symmetry \cite{Sheikh-Jabbari:2008re}. For any given QFT on commutative Minkowski space its VSR invariant counterpart is a noncommutative QFT, NCQFT, which is obtained by replacing the usual product of field operators with the nonlocal Moyal *-product. The NCQFT on noncommutative Moyal plane with light-like noncommutativity realization of VSR actually needs to twist deform the VSR subgroup Poincar\'e group.

Inspiring by the fact that Poincar\'e group admits the unique deformation into de Sitter group, Gibbons, Gomis  and Pope find that the subgroup ISIM(2) considered by Cohen and Glashow admits a 2-parameter family of continuous deformations which may be viewed as a quantum corrections or the quantum gravity effect to the very special relativity, but none of these give rise to noncommutative translations analogous to those of the de Sitter deformation of the Poincar\'e group: space-time remains flat. Among the 2-parameter family of deformation of $ISIM(2)$, they find that only a 1-parameter $DISIM_b (2)$, the deformation of $SIM(2)$, is physically acceptable \cite{Gibbons:2008re}. The line element invariant under $DISIM_b (2)$ is Lorentz violating and of Finsler type, $d{{s}^{2}}={{\left( {{\eta }_{\mu \upsilon }}d{{x}^{\mu }}d{{x}^{\upsilon }} \right)}^{1-b}}{{\left( {{n}_{\mu }}d{{x}^{\mu }} \right)}^{2b}}$. The $DISIM_b (2)$ invariant action for point particle and the wave equations for spin $0$, $\frac{1}{2}$ and $1$ are derived in their paper. The equation for spin $0$ field is in general a nonlocal equation, since it involves fractional even irrational derivatives. 

In our previous paper we follow Gibbons-Gomis-Pope's approach on the deformation of $ISIM(2)$ and investigate the deformation of all such kind of subgroups of Poincar\'e group which are the semi-product of three generators and four generators Lorentz subgroups with the spacetime translation group $T(4)$ (semi-product Poincar\'e subgroup) and the five dimensional representations, which are inherited from the five dimensional representation of Poincar\'e group, (the natural representation) of all the semi-product Poincar\'e subgroup as well as their deformed partners \cite{Zhang:2012ty}. We find that the deformation of semi-product Poincar\'e subgroup may have more than one families that are physically acceptable. There may be more than one inequivalent natural representations for one family of deformation of a specific Poincar\'e subgroup. Usually the deformation of the original Lorentz subgroup part causes the rotational operation an additional accompanied scale factor which is not reasonable for we believe that the departure from Lorentz symmetry should be from boost rather than rotational operation. Anyhow most deformed boost operations do indeed have an additional accompanied scale factors which will play a key role in the search of group action invariant geometry and construction of field theories in the spactime of the invariant geometry.

In the present paper, we investigate all the possible Finsler geometry realization of spacetime possessing the semi-product Poincar\'e subgroups and their deformed partner symmetry. To deal with the additional accompanied scale transformation of rotation and boost operation, we find all the independent scale covariant rank 1,2 and 3 tensors for all cases of the symmetry groups. The existence of invariant metric function automatically excludes the additional accompanied scale transformation of rotation operation which is consistent with our argument that the lorentz invariance should not be broken in rotation but in boost \cite{Zhang:2012ty}. We find that the admissible invariant metric function contains an arbitrary function of the specified combination of variables freedom, which can not be fixed by the investigation of symmetry. To fix the freedom, it is needed to study the dynamics in the corresponding spacetime. We investigate the dynamics of particles and field theories in our next subsequent pater.

This paper is organized as follows. We first give a brief introduction to Finsler geometry which leads the concept of Finsler-Minkowski spacetime which we concentrate in this paper. Then we give the general methods to construct the invariant metric function under some group action. The most part of the paper is devoted to the seeking of invariant metric functions under the groups we obtained in our previous paper \cite{Zhang:2012ty}. At last we give a brief discussion of our result and outlook of the investigation on dynamics in our next subsequent paper. We notice that there is also other approach indicating that VSR may be realized by Finsler geometry \cite{Li:2010iy}.

\section{Finsler Geometry}

Let us start with a brief review of the basic notions relevant for Finsler geometries \cite{Bao:2000ay,Shen:2001ly,Bao:2004ay,Chern:2005ry,Mo:2006ay,Dahl:2006ay}.

\subsection{The Metric Structure}
In Riemann geometry, the line element is of the form
\beq\label{eq:lineelement}
ds=\sqrt{{{g}_{\mu \upsilon }}d{{x}^{\mu }}d{{x}^{\upsilon }}}
\eeq
where ${{g}_{\mu \upsilon }}$ is the metric tensor of the manifold, which is the function of $x^{\mu}$. The Finsler geometry is a generalization of Riemamm geometry with the more general form of line element
\beq\label{eq:genlineelement}
ds=F\left (x^{\mu},dx^{\nu} \right ),
\eeq
where $F$ is defined on the tangent bundle of the manifold and a degree 1 homogenous function of $dx^{\mu}$, which includes the Riemann metric as a special case.  More specifically, the Finsler metric norm $F: TM \mapsto \mathbb{R}$ which is a real function $F(x,y)$ of a
spacetime point $x$ and of a tangent vector $y\in T_x M$, satisfies
\begin{itemize}
  \item Regularity: F is smooth on the entire slit tangent bundle $TM\backslash 0$,
  \item Non-degeneracy: $F(x,y)\neq 0$ if $ y \neq \mathbf{0}$,
  \item Positive homogeneity: $F(x,\lambda y)= |\lambda | F(x,y), \forall\lambda \in \mathbb{R}$.
\end{itemize}

The so called fundamental tensor can be defined
\begin{equation}\label{eq:finslermetric}
{{g}_{\mu \upsilon }}=\frac{\text{1}}{\text{2}}{{\dot{\partial }}_{\mu }}{{\dot{\partial }}_{\upsilon }}{{F}^{2}},
\end{equation}
where and hereafter ${{\hat{\partial }}_{\mu }}$ represents ${\partial }/{\partial {{x}^{\mu }}}\;$ and ${{\dot{\partial }}_{\mu }}$ represents ${\partial }/{\partial {{y}^{\mu }}}\;$, which is required to be continuous and non-degenerate.
It can be shown that \eqref{eq:finslermetric}
is equivalent to
\beq \label{eq:finslergmn}
F(x,y) = \sqrt{g_\mn(x,y)y^\mu y^\nu},
\eeq
which shows that $g_\mn(x,y)$ is a homogeneous function of degree
zero of the vector $y$.  Also, since by definition $g_\mn$ is non degenerate, it admits an inverse $g^\mn$  such that $ g^\mn(x,y)g_{\nu\alpha}(x,y)={\delta^{\mu}}_\alpha$, a metric tensor similar to one in Riemann geometry with the difference from Riemann geometry that the metric tensor here does not depend only on coordinates of base manifold but also coordinates of tangent space.

The Finsler metric tensor thus defined must be index symmetric. The derivatives to $y^{\mu}$ are also index symmetric,
\beq\label{eq:symmderv}
{{\dot{\partial }}_{\alpha }}{{g}_{\mu \upsilon }}={{\dot{\partial }}_{\mu }}{{g}_{\upsilon \alpha }}={{\dot{\partial }}_{\upsilon }}{{g}_{\alpha \mu }}.
\eeq
%We can build various geometric quantities like the connection, Riemann curvature tensor %and Ricci curvature tensor etc. with the metric tensor similar to Riemann geometry.
One can introduce the Christoffel symbols of the first and second kind in terms of Finsler metric tensor. There is a connection between the so called spray induced by $F$ and the Christoffel symbols,
\beq\label{eq:spray}
G{\left( y \right)^\mu } = \frac{1}{2}\gamma _{\alpha \beta }^\mu {y^\alpha }{y^\beta },
\eeq
where
\beq\label{eq:tangent}
\gamma _{\alpha \beta }^{\mu }=\frac{1}{2}{{g}^{\mu \upsilon }}\left( {{{\hat{\partial }}}_{\alpha }}{{g}_{\beta \upsilon }}+{{{\hat{\partial }}}_{\beta }}{{g}_{\alpha \upsilon }}-{{{\hat{\partial }}}_{\upsilon }}{{g}_{\alpha \beta }} \right)
\eeq
are the formal Christoffel symbols of the second kind.

Apart from the well-known geometric quantities in Riemann geometry, there are many geometric quantities which are non-Riemannian and unique for Finsler geometry, e.g. the Cartan tensor which specifies the departure of the manifold from Riemannian
\beq\label{eq:cartantensor}
{C_{\alpha \beta \sigma }} = \frac{1}{2}{\dot{\partial }}_\sigma{g_{\alpha \beta }}.
\eeq
It is apparent that the Cartan tensor is full symmetric. The manifold with zero Cartan tensor is Riemannian and viceversa. The length of Cartan co-vector ${C_\mu } = {C_{\mu \alpha \beta }}{g^{\alpha \beta }}$, $C = {g^{\mu \upsilon }}{C_\mu }{C_\upsilon }$, can be utilized to describe the departure of a Finsler manifold from Riemannian.

Finsler geometry can be regarded as the geometry of the tangent bundle $TM$. The local coordinate $x^{\mu}$ of $x \in M$ give rise to the local coordinate of $\{ x^{\mu}, y^{\alpha} \} \in TM$ through the mechanism
\beq\label{eq:tangent}
y=y^\mu \frac{\partial}{\partial x^\mu}
\eeq
where  ${{y}^{\mu }}={{y}^{\mu }}\left( {{x}^{\alpha }} \right)$ are fiberwise global. Under local coordinate transformation $x{{'}^{\mu }}=x{{'}^{\mu }}\left( x \right)$, the vector of tangent space transforms like
\beq\label{eq:transtangent}
y{{'}^{\mu }}=\frac{\partial x{{'}^{\mu }}}{\partial {{x}^{\alpha }}}{{y}^{\alpha }}.
\eeq
So under coordinate transformation
\beq\label{eq:tangent}
\left\{ \begin{aligned}
  & x{{'}^{\mu }}=x{{'}^{\mu }}\left( x \right) ,\\
 & y{{'}^{\mu }}=y{{'}^{\mu }}\left( x,y \right) ,
\end{aligned}\right.
\eeq
the coordinate base vectors transform as
\beq\label{eq:cotrans}
\left\{ \begin{aligned}
  & \hat{\partial }{{'}_{\mu }}=\frac{\partial {{x}^{\alpha }}}{\partial x{{'}^{\mu }}}{{{\hat{\partial }}}_{\alpha }}+\frac{{{\partial }^{2}}{{x}^{\alpha }}}{\partial x{{'}^{\mu }}\partial x{{'}^{\upsilon }}}y{{'}^{\upsilon }}{{{\dot{\partial }}}_{\alpha }} ,\\
 & \dot{\partial }{{'}_{\mu }}=\frac{\partial {{x}^{\alpha }}}{\partial x{{'}^{\mu }}}{{{\dot{\partial }}}_{\alpha }} .
\end{aligned}\right.
\eeq
The tangent bundle of the manifold $TM$ has a local coordinate basis
that consists of the $\frac{\partial}{\partial x^{\mu}}$ and the $\frac{\partial}{\partial y^{\alpha}}$ . However, under the transformation on $TM$ induced by a coordinate change on M, the vectors
$\frac{\partial}{\partial x^{\mu}}$ transform in a somewhat complicated manner, while the $\frac{\partial}{\partial y^{\alpha}}$ do not have this problem. The cotangent bundle of $TM$ has a local coordinate basis $\left\{dx^{\mu}, dy^{\alpha}\right\}$. Under the induced transformation, the behavior of $dx^{\mu}$ is simple while one of $dy^{\alpha}$ is not.

The remedy lies in replacing $\frac{\partial}{\partial x^{\mu}}$ by
\beq\label{eq:vecbaschg}
\frac{\delta}{\delta x^{\mu}}=\frac{\partial}{\partial x^{\mu}}-N^{\alpha}_{\mu} \frac{\partial}{\partial y^{\alpha}}
\eeq
and $dy^{\alpha}$ by
\beq\label{eq:ybasechg}
\delta y^{\alpha}=dy^{\alpha}+N^{\alpha}_{\mu}dx^{\mu},
\eeq
where
\beq\label{eq:nonlconn}
 N_{\mu }^{\upsilon }=\frac{1}{2}{{y}^{\alpha }}{{y}^{\beta }}{{\dot{\partial }}_{\mu }}\gamma _{\alpha \beta }^{\upsilon }+\gamma _{\mu \alpha }^{\upsilon }{{y}^{\alpha }}
\eeq
are the Finsler nonlinear connection.

They indeed have simple behavior under transformations induced
by coordinate changes on $M$. Thus the two new natural(local) bases that are dual to each other are
\begin{itemize}
  \item $\{\frac{\delta}{\delta x^{\mu}}, F\frac{\partial}{\partial y^{\alpha}}\}$ for the tangent bundel of $TM \setminus {\bf 0}$,
  \item $\{dx^{\mu}, \delta y^{\alpha}\}$ for the tangent bundel of $TM \setminus {\bf 0}$.
\end{itemize}

Moreover there is a relation between the non-linear connection and the spray
\beq\label{eq:spray}
N_\upsilon ^\mu  = {\dot{\partial} _\upsilon }{G^\mu }.
\eeq

\subsection{The Connection Structure}
In Finsler geometry, the covariant derivative need to be carried on two sets of coordinates, so the connection structure is lager than Riemann geometry,
\beq\label{eq:connfinsler}
\left\{ \begin{array}{l}
{\hat\nabla _\mu }{A^\alpha } = {\hat\partial _\mu }{A^\alpha } + \Gamma _{\beta \mu }^\alpha {A^\beta }, \\
{\hat\nabla _\mu }{A_\alpha } = {\hat\partial _\mu }{A_\alpha } - \Gamma _{\alpha \mu }^\beta {A_\beta },\\
{{\dot{\nabla} }_\mu }{A^\alpha } = {{\dot{\partial} }_\mu }{A^\alpha } + \Lambda _{\beta \mu }^\alpha {A^\beta }, \\{{\dot{\nabla} }_\mu }{A_\alpha } = {{\dot{\partial} }_\mu }{A_\alpha } - \Lambda _{\alpha \mu }^\beta {A_\beta }.
\end{array} \right.
\eeq

Similar to Riemann geometry, one can impose the adaptable condition between metric and connection by
\beq\label{eq:adaption}
{\hat\nabla _\mu }{g_{\alpha \beta }} = \dot{ \nabla} {_\mu }{g_{\alpha \beta }} = 0.
\eeq
One has
\beq\label{eq:adaptconn}
\left\{ \begin{array}{l}
\Gamma _{\mu \upsilon }^\sigma  = \frac{1}{2}{g^{\sigma \rho }}\left( {{\delta _\mu }{g_{\rho \upsilon }} + {\delta _\upsilon }{g_{\rho \mu }} - {\delta _\rho }{g_{\mu \upsilon }}} \right),\\
\Lambda _{\mu \upsilon }^\sigma  = \frac{1}{2}{g^{\sigma \rho }}{{\dot{\partial} }_\rho }{g_{\mu \upsilon }},
\end{array} \right.
\eeq
where $\Gamma _{\mu \upsilon }^\sigma $ is called the Chern connection and the adaptable connection in tangent space is just the Cartan tensor. The triple $\left( {N_\mu ^\upsilon ,\Gamma _{\mu \upsilon }^\sigma ,\Lambda _{\mu \upsilon }^\sigma } \right)$ is called Cartan connection. The other three popular Finsler connections are
\begin{enumerate}
  \item Chern-Rund connection $\left( {N_\mu ^\upsilon ,\Gamma _{\alpha \beta }^\mu ,0} \right)$,
  \item Berwald connection $\left( {N_\mu ^\upsilon ,G_{\alpha \beta }^\mu ,0} \right)$,
  \item Hashiguchi connection $\left( {N_\mu ^\upsilon ,G_{\mu \upsilon }^\sigma ,C_{\mu \upsilon }^\sigma } \right)$,
\end{enumerate}
where $G_{\mu \upsilon }^\sigma $ is called Berwald connection defined by $G_{\mu \upsilon }^\sigma  = {\dot{\partial} _\upsilon }N_\mu ^\sigma  = {\dot{\partial} _\mu }{\dot \partial _\upsilon }{G^\sigma }$. Berwald connection and Chern connection are connected each other by a full symmetric Landsberg tensor ${L_{\alpha \beta \sigma }}$,
\beq\label{eq:bewchern}
G_{\mu \upsilon }^\sigma  = \Gamma _{\mu \upsilon }^\sigma  + L_{\mu \upsilon }^\sigma .
\eeq

In Riemann geometry, torsion is free from the adaption condition of connection while in Finsler geometry, torsion is somewhat inevitable,
\beq\label{eq:torsion}
\left\{ \begin{array}{l}
\left[ {{\delta _\mu },{\delta _\upsilon }} \right] = R_{\mu \upsilon }^\alpha {{\dot \partial }_\alpha },\\
\left[ {{\delta _\mu },{{\dot \partial }_\upsilon }} \right] = G_{\mu \upsilon }^\alpha {{\hat \partial }_\alpha },\\
\left[ {{{\dot \partial }_\mu },{{\dot \partial }_\upsilon }} \right] = 0,
\end{array} \right.
\eeq
where $R_{\mu \upsilon }^\sigma $ is the Finsler torsion tensor
\beq\label{eq:torexp}
R_{\mu \upsilon }^\sigma  = {\delta _\upsilon }N_\mu ^\sigma  - {\delta _\mu }N_\upsilon ^\sigma ,
\eeq
which does not vanish in Riemann geometry and is just $R_{\mu \upsilon }^\sigma  = {y^\alpha }R_{\alpha \mu \upsilon }^\sigma $, where $R_{\alpha \mu \upsilon }^\sigma $ is the Riemann curvature tensor.

\subsection{Curvature}
In Finsler geometry, there are several tensors which can be used to describe the curvature of the manifold.

The Finsler curvature tensor is
\beq\label{eq:fincur}
F_{\sigma \mu \upsilon }^\rho  = {\dot \partial _\sigma }R_{\mu \upsilon }^\rho ,
\eeq
which is just the Riemann curvature tensor in Riemann geometry.

The flag curvature tensor is
\beq\label{eq:flagcur}
F_\mu ^\upsilon  = {y^\alpha }R_{\mu \alpha }^\upsilon .
\eeq

The non-Riemannian Berwald curvature tensor is
\beq\label{eq:berwcur}
B_{\rho \mu \upsilon }^\sigma  = {\dot \partial _\rho }G_{\mu \upsilon }^\sigma .
\eeq
The Berwald connection leads to another rank 4 curvature tensor
\beq\label{eq:ricciwcur}
\tilde R_{\rho \mu \upsilon }^\sigma  = {\delta _\mu }G_{\rho \upsilon }^\sigma  - {\delta _\upsilon }G_{\rho \mu }^\sigma  + G_{\rho \upsilon }^\gamma G_{\gamma \mu }^\sigma  - G_{\rho \mu }^\gamma G_{\gamma \upsilon }^\sigma.
\eeq

The analogous curvature tensor of Riemann geometry in a Finsler setting for connection $\left( {N_\mu ^\upsilon ,H_{\alpha \beta }^\mu ,V_{\alpha \beta }^\mu } \right)$ is
\beq\label{eq:anavur}
\left\{ \begin{array}{l}
S_{\sigma \alpha \beta }^\mu  = {\delta _\beta }H_{\sigma \alpha }^\mu  - {\delta _\alpha }H_{\sigma \beta }^\mu  + H_{\rho \beta }^\mu H_{\sigma \alpha }^\rho  - H_{\rho \alpha }^\mu H_{\sigma \beta }^\rho ,\\
P_{\sigma \alpha \beta }^\mu  = {{\dot \partial }_\beta }V_{\sigma \alpha }^\mu  - {{\dot \partial }_\alpha }V_{\sigma \beta }^\mu  + V_{\rho \beta }^\mu V_{\sigma \alpha }^\rho  - V_{\rho \alpha }^\mu V_{\sigma \beta }^\rho ,\\
X_{\sigma \alpha \beta }^\mu  = {{\dot \partial }_\beta }H_{\sigma \alpha }^\mu  - {\delta _\alpha }V_{\sigma \beta }^\mu  + V_{\rho \beta }^\mu H_{\sigma \alpha }^\rho  - H_{\rho \beta }^\mu V_{\sigma \alpha }^\rho,
\end{array} \right.
\eeq
where $S_{\sigma \alpha \beta }^\mu $ is the curvature tensor in the pure horizontal direction, $P_{\sigma \alpha \beta }^\mu $ is the curvature tensor in the pure vertical direction while $X_{\sigma \alpha \beta }^\mu $ is the curvature tensor in the mixed directions.

There is a special class of Finsler manifold which is worthy to pay attention, the Minkowski manifold. In Finsler geometry, Minkowski manifold is a class of flat manifolds whose Finsler norm does not change with the coordinate on the base manifold and hence a function of the coordinate of the vector space, $F=F ( y^{\alpha})$. The metric tensor depends only on $ y^{\alpha}$ too. So $\gamma _{\alpha \beta }^{\mu }=0$ and the non-linear connection $N_{\mu }^{\upsilon }=0$ which lead all the connections of Minkowski manifold to be zero and the zero curvatures. So Finsler-Minkowski manifold is flat. In this paper, we concentrate on the Finsler-Minkowski manifold to seek the invariant metric under semi-direct product Poincar\'e subgroups and their deformed partners.

\section{The invariant metric of spacetime under semi-direct product Poincar\'e subgroups and their deformed partners}
In principle, the spacetime geometry does not have to be Riemann geometry. The reason why it is not some more general type of geometry, e.g. Finsler geometry, but Riemannian in the scheme of general relativity is because of equivalence principle when the local symmetry is Lorentz group. In the scheme of VSR, the local symmetry is not the entire Lorentz group but its proper subgroup. It is not necessary that the general very special relativity has to be Riemannian any more. In fact it is revealed that the general very special relativity is Finslerian under a special deformation $Disim_b(2)$ of $Isim(2)$ symmetry. We have systematically investigate all the possible deformation of Poincar\'e subalgebra which include all the possible symmetry of very spacial relativity and get their corresponding natural matrix representations already. In this section we will investigate what kind of Finsler geometries that all the possible deformation of Poincar\'e subalgebra correspond to.

Without losing generality, we assume that the Finlerian metric $F^2$ consists of $M$ parts factors,
\beq \label{eq:finmtr}
{F^2} = \prod\limits_{i = 1}^M {{F_i}} .
\eeq
The $F_i$ has the form
\beq \label{eq:finmtrpart}
{F_i} = M_i^{{E_i}} ,
\eeq
where $E_i$ is constant and $M_i$ satisfies
\beq \label{eq:facfinmtr}
{M_i}\left( {{y^\mu }} \right) = {G_{{\mu _1}{\mu _2}...{\mu _{{p_i}}}}}\prod\limits_{j = 1}^{{p_i}} {{y^{{\mu _j}}}} .
\eeq
The $G_{{\mu _1}{\mu _2}...{\mu _{{p_i}}}}$ is constant tensor. So $F_i$ is a degree $p_iE_i$ homogenous function of tangent space coordinates $y_{\mu}$. For $F^2$ is a degree 2 homogenous function of $y_{\mu}$, we have
\beq \label{eq:powercnstr}
\sum\limits_{i = 1}^M {{p_i}{E_i}}  = 2 .
\eeq
Suppose $T_a$ is the group element of single parameter Lie group generated by the generator of spacetime symmetry group, we can demand that under the action of $T_a$, $M_i$ satisfies
\beq \label{eq:miunderta}
{M_i}\left( {{T_a}\left( {{y^\mu }} \right)} \right) = {A_{ia}}{M_i}\left( {{y^\mu }} \right).
\eeq
Suppose the action of $T_a$ on the coordinates $x^{\mu}$ of spacetime manifold is
\beq \label{eq:taonx}
{T_a}\left( {{x^\mu }} \right) = \left( {{R_a}} \right)_\alpha ^\mu {x^\alpha } + {\left( {{P_a}} \right)^\mu },
\eeq
then the action of $T_a$ on the coordinates $y^{\mu}$ of the tangent space is
\beq \label{eq:taony}
{T_a}\left( {{y^\mu }} \right) = \left( {{R_a}} \right)_\alpha ^\mu {y^\alpha }.
\eeq

From \eqref{eq:facfinmtr} and \eqref{eq:miunderta}, we have
\beq \label{eq:rarel}
\prod\limits_{j = 1}^{{p_i}} {\left( {{R_a}} \right)_{{\mu _j}}^{{\alpha _j}}} {G_{{\alpha _1}{\alpha _2}...{\alpha _{{p_i}}}}} = {A_{ia}}{G_{{\mu _1}{\mu _2}...{\mu _{{p_i}}}}}.
\eeq
For $F^2$ is invariant under the action of $T_a$, we have
\beq \label{eq:invmetrscalefactor}
\prod\limits_{i = 1}^M {A_{ia}^{{E_i}}}  = 1,
\eeq

For infinitesimal symmetric operation $\left( {{R_a}} \right)_\alpha ^\mu =\delta _\alpha ^\mu  + \theta {\phi _a}_\alpha ^\mu $, we have
\beq \label{eq:infaction}
\sum\limits_{j = 1}^{{p_i}} {{\phi _a}_{{\mu _j}}^{{\alpha _j}}{G_{{\mu _1}...{\alpha _j}...{\mu _{{p_i}}}}}}  = A{'_{ia}}{G_{{\mu _1}{\mu _2}...{\mu _{{p_i}}}}},
\eeq
where $A{'_{ia}} = {\left. {\frac{{d{A_{ia}}}}{{d\theta }}} \right|_{\theta  = 0}}$.

We then can find out all of ${G_{{\mu _1}...{\mu _{{p_i}}}}}$ which satisfy \eqref{eq:infaction} and construct reasonable invariant metric via \eqref{eq:invmetrscalefactor}.
\subsection{The de Sitter group}

The deformed group of Poincar\'e group is de Sitter group. As in \cite{Zhang:2012ty}, its natural representation is

\beq \label{eq:reptrands}
\begin{array}{l}
{p_t} = \left( {\begin{array}{*{20}{c}}
{}&{}&{}&{}&1\\
{}&{}&{}&0&{}\\
{}&{}&0&{}&{}\\
{}&0&{}&{}&{}\\
{ - \lambda }&{}&{}&{}&{}
\end{array}} \right),
{p_x} = \left( {\begin{array}{*{20}{c}}
0&{}&{}&{}&{}\\
{}&0&{}&{}&1\\
{}&{}&0&{}&{}\\
{}&{}&{}&0&{}\\
{}&\lambda &{}&{}&0
\end{array}} \right),\\
{p_y} = \left( {\begin{array}{*{20}{c}}
0&{}&{}&{}&{}\\
{}&0&{}&{}&{}\\
{}&{}&0&{}&1\\
{}&{}&{}&0&{}\\
{}&{}&\lambda &{}&0
\end{array}} \right),
{p_z} = \left( {\begin{array}{*{20}{c}}
0&{}&{}&{}&{}\\
{}&0&{}&{}&{}\\
{}&{}&0&{}&1\\
{}&{}&{}&0&{}\\
{}&{}&\lambda &{}&0
\end{array}} \right).
\end{array}
\eeq
The group elements of single parameter Lie group generated by the corresponding generator are
\beq \label{eq:grpeleds}
\begin{array}{l}
{P_t}\left( \theta  \right) = \left( {\begin{array}{*{20}{c}}
{\cos \left( {\theta \sqrt \lambda  } \right)}&{}&{}&{}&{\frac{{\sin \left( {\theta \sqrt \lambda  } \right)}}{{\sqrt \lambda  }}}\\
{}&1&{}&{}&{}\\
{}&{}&1&{}&{}\\
{}&{}&{}&1&{}\\
{ - \sqrt \lambda  \sin \left( {\theta \sqrt \lambda  } \right)}&{}&{}&{}&{\cos \left( {\theta \sqrt \lambda  } \right)}
\end{array}} \right),\\
{P_x}\left( \theta  \right) = \left( {\begin{array}{*{20}{c}}
1&{}&{}&{}&{}\\
{}&{\cosh \left( {\theta \sqrt \lambda  } \right)}&{}&{}&{\frac{{\sinh \left( {\theta \sqrt \lambda  } \right)}}{{\sqrt \lambda  }}}\\
{}&{}&1&{}&{}\\
{}&{}&{}&1&{}\\
{}&{\sqrt \lambda  \sinh \left( {\theta \sqrt \lambda  } \right)}&{}&{}&{\cosh \left( {\theta \sqrt \lambda  } \right)}
\end{array}} \right),\\
{P_y}\left( \theta  \right) = \left( {\begin{array}{*{20}{c}}
1&{}&{}&{}&{}\\
{}&1&{}&{}&{}\\
{}&{}&{\cosh \left( {\theta \sqrt \lambda  } \right)}&{}&{\frac{{\sinh \left( {\theta \sqrt \lambda  } \right)}}{{\sqrt \lambda  }}}\\
{}&{}&{}&1&{}\\
{}&{}&{\sqrt \lambda  \sinh \left( {\theta \sqrt \lambda  } \right)}&{}&{\cosh \left( {\theta \sqrt \lambda  } \right)}
\end{array}} \right),\\
{P_z}\left( \theta  \right) = \left( {\begin{array}{*{20}{c}}
1&{}&{}&{}&{}\\
{}&1&{}&{}&{}\\
{}&{}&1&{}&{}\\
{}&{}&{}&{\cosh \left( {\theta \sqrt \lambda  } \right)}&{\frac{{\sinh \left( {\theta \sqrt \lambda  } \right)}}{{\sqrt \lambda  }}}\\
{}&{}&{}&{\sqrt \lambda  \sinh \left( {\theta \sqrt \lambda  } \right)}&{\cosh \left( {\theta \sqrt \lambda  } \right)}
\end{array}} \right).
\end{array}
\eeq
Note that the representation is inherited from the natural representation of Poincar\'e group in which the representation space has a natural meaning of spacetime and the matrices have the features that the upper left $4\times 4$ part of represents rotation and boost, the upper right $1\times 4$ part represents translation and the lower $5\times 1$ part should keep to be zero. Matrices in \eqref{eq:reptrands} indicate that the de Sitter invariant spacetime must be a curved space and the invariant metric is expected to be coordinate dependent. The direct search by \eqref{eq:miunderta} or \eqref{eq:infaction} shows that there are invariant tensor neither of rank one or two nor three or four, i.e. the de Sitter invariant metric can not satisfies \eqref{eq:minkmetric}, which is only satisfied by Minkowski manifold.

\subsection{The $DISIM$ group}
There are two subclasses for $DISIM$, one denoted by $DISIM$ in which the $SIM$ part is undeformed, while the other in which the $SIM$ part is deformed and can be further specified into two different deformation group $XDISIM1$ and $XDISIM2$.
\subsubsection{The $DISIM$ group}
The deformed generators in $disim$ are $r_z$ and $b_z$ with the natural representation matrices,
\beq \label{eq:rzberepmat}
{r_z} = \left( {\begin{array}{*{20}{c}}
{{A_1}}&{}&{}&{}&{}\\
{}&{{A_1}}&{ - 1}&{}&{}\\
{}&1&{{A_1}}&{}&{}\\
{}&{}&{}&{{A_1}}&{}\\
{}&{}&{}&{}&0
\end{array}} \right),{b_z} = \left( {\begin{array}{*{20}{c}}
{{A_2}}&{}&{}&1&{}\\
{}&{{A_2}}&{}&{}&{}\\
{}&{}&{{A_2}}&{}&{}\\
1&{}&{}&{{A_2}}&{}\\
{}&{}&{}&{}&0
\end{array}} \right).
\eeq
The corresponding one parameter group elements are,
\beq \label{eq:rzbegrpele}
\begin{array}{l}
{R_z}\left( \theta  \right) = {e^{{A_1}\theta }}\left( {\begin{array}{*{20}{c}}
1&{}&{}&{}&{}\\
{}&{\cos \theta }&{ - \sin \theta }&{}&{}\\
{}&{\sin \theta }&{\cos \theta }&{}&{}\\
{}&{}&{}&1&{}\\
{}&{}&{}&{}&{{e^{ - {A_1}\theta }}}
\end{array}} \right),\\
{B_z}\left( \theta  \right) = {e^{{A_2}\theta }}\left( {\begin{array}{*{20}{c}}
{\cosh \theta }&{}&{}&{\sinh \theta }&{}\\
{}&1&{}&{}&{}\\
{}&{}&1&{}&{}\\
{\sinh \theta }&{}&{}&{\cosh \theta }&{}\\
{}&{}&{}&{}&{{e^{ - {A_2}\theta }}}
\end{array}} \right).
\end{array}
\eeq
There exists neither rank 1 nor rank 2 invariant tensor. However, there are indeed conformal covariant rank 1 and rank 2 tensors, e.g. the rank 1 tensor, known as spurion,
\beq \label{eq:spurion}
{N_\mu } = \left( {\begin{array}{*{20}{c}}
1\\
0\\
0\\
1
\end{array}} \right),
\eeq
which transforms conformally as ${R_z}\left( \theta  \right)\left( {{N_\mu }} \right) = {e^{{A_1}\theta }}{N_\mu }$ under $R_z(\theta)$ and as ${B_z}\left( \theta  \right)\left( {{N_\mu }} \right) = {e^{\left( {1 + {A_2}} \right)\theta }}{N_\mu }$ under ${B_z}\left( \theta  \right)$.

The conformally covariant rank 2 tensor under the action of $DISIM$ is the Minkowski metric tensor,
\beq \label{eq:minkmtrtns}
{G_{\mu \upsilon }} = \left( {\begin{array}{*{20}{c}}
{ - 1}&{}&{}&{}\\
{}&1&{}&{}\\
{}&{}&1&{}\\
{}&{}&{}&1
\end{array}} \right),
\eeq
which transforms as ${R_z}\left( \theta  \right)\left( {{G_{\mu \upsilon }}} \right) = {e^{2{A_1}\theta }}{G_{\mu \upsilon }}$ under ${R_z}\left( \theta  \right)$ and as ${B_z}\left( \theta  \right)\left( {{G_{\mu \upsilon }}} \right) = {e^{2{A_2}\theta }}{G_{\mu \upsilon }}$ under ${B_z}\left( \theta  \right)$.

The rank 3 conformally covariant tensor has the form,
\beq \label{eq:rank3tensor}
\begin{array}{l}
{F_{t\mu \upsilon }} = \left( {\begin{array}{*{20}{c}}
3&{}&{}&1\\
{}&{ - 1}&{}&{}\\
{}&{}&{ - 1}&{}\\
1&{}&{}&{ - 1}
\end{array}} \right),
{F_{x\mu \upsilon }} = \left( {\begin{array}{*{20}{c}}
{}&{ - 1}&{}&{}\\
{ - 1}&{}&{}&{ - 1}\\
{}&{}&{}&{}\\
{}&{ - 1}&{}&{}
\end{array}} \right),\\
{F_{y\mu \upsilon }} = \left( {\begin{array}{*{20}{c}}
{}&{}&{ - 1}&{}\\
{}&{}&{}&{}\\
{ - 1}&{}&{}&{ - 1}\\
{}&{}&{ - 1}&{}
\end{array}} \right),
{F_{z\mu \upsilon }} = \left( {\begin{array}{*{20}{c}}
1&{}&{}&{ - 1}\\
{}&{ - 1}&{}&{}\\
{}&{}&{ - 1}&{}\\
{ - 1}&{}&{}&{ - 3}
\end{array}} \right),
\end{array}
\eeq
which has the conformal factor ${e^{3{A_1}\theta }}$ under ${R_z}\left( \theta  \right)$ and ${e^{\left( {1 + 3{A_2}} \right)\theta }}$ under ${B_z}\left( \theta  \right)$. Actually it is not an independent tensor and can be written as ${F_{\sigma \mu \upsilon }} = {N_{\left( \sigma  \right.}}{G_{\left. {\mu \upsilon } \right)}}$.

The invariant metric is therefore of the form,
\beq \label{eq:invmtrdisim}
{F^2} = {\left( {{N_\mu }{y^\mu }} \right)^A}{\left( {{G_{\mu \upsilon }}{y^\mu }{y^\upsilon }} \right)^B}.
\eeq
That the metric function $F^2$ is a degree 2 homogenous function of $y_{\mu}$ and the invariance under the action of $DISIM$, esp. under ${R_z}\left( \theta  \right)$ and under ${B_z}\left( \theta  \right)$, gives the constrain condition,
\beq \label{eq:cstronmtr}
\left\{ \begin{array}{l}
A + 2B = 2\\
A{A_1}\theta  + 2B{A_1}\theta  = 0\\
A\left( {1 + {A_2}} \right)\theta  + 2B{A_2}\theta  = 0
\end{array} \right.,
\eeq
where the first one comes from $F^2$ as a degree 2 homogenous function, the second one from invariance under ${R_z}\left( \theta  \right)$ and the third from ${B_z}\left( \theta  \right)$ respectively. The first and the second constrain gives $A_1=0$. It means that there does not exist deformed $R_z$ invariant Minkowski-Finsler type of spacetime. Among $DISIM$ groups, only those in which $R_z$ is not deformed and only $B_z$ is deformed, denoted by $DISIMb$ have the Minkowski-Finsler type of invariant metric. The constrain \eqref{eq:cstronmtr} then becomes,
\beq \label{eq:cstrwithbz}
\left\{ \begin{array}{l}
A + 2B = 2\\
A\left( {1 + {A_2}} \right)\theta  + 2B{A_2}\theta  = 0
\end{array} \right.,
\eeq
which has the solution
\beq \label{eq:solnofcstr}
\left\{ \begin{array}{l}
A =  - 2{A_2}\\
B = 1 + {A_2}
\end{array} \right.
\eeq
The $DISIMb$ invariant metric function is
\beq \label{eq:disimbmetric}
{F^2} = {\left( {{N_\mu }{y^\mu }} \right)^{ - 2{A_2}}}{\left( {{G_{\mu \upsilon }}{y^\mu }{y^\upsilon }} \right)^{1 + {A_2}}},
\eeq
where $A_2$ is a free parameter which parametrizes the $DISIMb$ group.

\subsubsection{$XDISIM1$ and $XDISIM2$ groups}

We are going to find the invariant metric for $XDISIM1$ and $XDISIM2$ groups.

The deformed generators in $XDISIM1$ are,
\beq \label{eq:dfmgnrxdisim1}
\begin{array}{l}
{r_z} = \left( {\begin{array}{*{20}{c}}
{{A_2}}&{}&{}&{}&{}\\
{}&{{A_2}}&{ - 1}&{}&{}\\
{}&1&{{A_2}}&{}&{}\\
{}&{}&{}&{{A_2}}&{}\\
{}&{}&{}&{}&0
\end{array}} \right),\\
{b_z} = \left( {\begin{array}{*{20}{c}}
{{A_3} - {A_1}}&{}&{}&{1 + {A_1}}&{}\\
{}&{{A_3} - {A_1}}&{}&{}&{}\\
{}&{}&{{A_3} - {A_1}}&{}&{}\\
{1 + {A_1}}&{}&{}&{{A_3} - {A_1}}&{}\\
{}&{}&{}&{}&0
\end{array}} \right),\\
{p_t} = \left( {\begin{array}{*{20}{c}}
0&{}&{}&{}&{1 + \frac{{{A_1}}}{2}}\\
{}&0&{}&{}&{}\\
{}&{}&0&{}&{}\\
{}&{}&{}&0&{\frac{{{A_1}}}{2}}\\
{}&{}&{}&{}&0
\end{array}} \right),
{p_z} = \left( {\begin{array}{*{20}{c}}
0&{}&{}&{}&{ - \frac{{{A_1}}}{2}}\\
{}&0&{}&{}&{}\\
{}&{}&0&{}&{}\\
{}&{}&{}&0&{1 + \frac{{3{A_1}}}{2}}\\
{}&{}&{}&{}&0
\end{array}} \right),\\
{p_x} = \left( {\begin{array}{*{20}{c}}
0&{}&{}&{}&{}\\
{}&0&{}&{}&{1 + {A_1}}\\
{}&{}&0&{}&{}\\
{}&{}&{}&0&{}\\
{}&{}&{}&{}&0
\end{array}} \right),
{p_y} = \left( {\begin{array}{*{20}{c}}
0&{}&{}&{}&{}\\
{}&0&{}&{}&{}\\
{}&{}&0&{}&{1 + {A_1}}\\
{}&{}&{}&0&{}\\
{}&{}&{}&{}&0
\end{array}} \right).
\end{array}
\eeq
The conformally covariant rank 1 tensor under the action of $XDISIM1$ is still $N_{\mu}$, which has conformal factor ${e^{{A_2}\theta }}$ and ${e^{\left( {1 + {A_3}} \right)\theta }}$ under $R_z(\theta)$ and $B_z(\theta)$ respectively. The conformally covariant rank 2 tensor is still Minkowski metric tensor $G_{\mu\nu}$, which has conformal factor ${e^{2{A_2}\theta }}$ and ${e^{2\left( {1 + {A_3}} \right)\theta }}$ respectively. So is the case for rank 3 tensor. The metric function is therefore
\beq \label{eq:mtrxdisim1}
{F^2} = {\left( {{N_\mu }{y^\mu }} \right)^{2\frac{{{A_1} - {A_3}}}{{1 + {A_1}}}}}{\left( {{G_{\mu \upsilon }}{y^\mu }{y^\upsilon }} \right)^{\frac{{1 + {A_3}}}{{1 + {A_1}}}}},
\eeq
which is apparently regresses to the form of $DISIM$ when $A_1=0$.

The deformed generators for $XDISIM2$ are,
\beq \label{eq:dfmgenxdisim2}
\begin{array}{c}
  {r_z} = \left( {\begin{array}{*{20}{c}}
{{A_2}}&{}&{}&{}&{}\\
{}&{{A_2}}&{ - 1}&{}&{}\\
{}&1&{{A_2}}&{}&{}\\
{}&{}&{}&{{A_2}}&{}\\
{}&{}&{}&{}&0
\end{array}} \right),\\
{b_z} = \left( {\begin{array}{*{20}{c}}
0&{}&{}&{1 + 2{A_1} - {A_3}}&{}\\
{}&{{A_3} - {A_1}}&{}&{}&{}\\
{}&{}&{{A_3} - {A_1}}&{}&{}\\
{1 + {A_3}}&{}&{}&{2\left( {{A_3} - {A_1}} \right)}&{}\\
{}&{}&{}&{}&0
\end{array}} \right) .
\end{array}
\eeq
The spurion $N_{\mu}$ is the conformal covariant rank 1 tensor under $XDISIM2$, with conformal factors ${e^{{A_2}\theta }}$ and ${e^{\left( {{A_3}-A_1} \right)\theta }}$ under $R_z(\theta)$ and $B_z(\theta)$ respectively. The rank 2 conformal covariant tensor has a form rather than $G_{\mu\nu}$ but
\beq \label{eq:hforxdisim2}
{H_{\mu \upsilon }} = \left( {\begin{array}{*{20}{c}}
{ - \frac{{1 + {A_3}}}{{1 + {A_1}}}}&{}&{}&{\frac{{{A_1} - {A_3}}}{{1 + {A_1}}}}\\
{}&1&{}&{}\\
{}&{}&1&{}\\
{\frac{{{A_1} - {A_3}}}{{1 + {A_1}}}}&{}&{}&{\frac{{1 + 2{A_1} - {A_3}}}{{1 + {A_1}}}}
\end{array}} \right),
\eeq
with conformal factors ${e^{2{A_2}\theta }}$ and ${e^{2\left( {A_3- {A_1}} \right)\theta }}$ under $R_z(\theta)$ and $B_z(\theta)$ respectively. What is interesting is that $H_{\mu\nu}$ can not return to $G_{\mu\nu}$ when $A_1=0$ but $G_{\mu\nu}$ with a coordinate transformation. Note that the spurion $N_{\mu}$ is still light like with $H_{\mu\nu}$ as the metric tensor,
\beq \label{eq:nhlightlike}
\left( {{N_\mu },{N_\upsilon }} \right) = {H^{\mu \upsilon }}{N_\mu }{N_\upsilon } = 0.
\eeq

The coordinate
\beq \label{eq:cdtrsfdiagh}
\left\{ \begin{array}{l}
t' = t\\
x' = x\\
y' = y\\
z' = z - t\frac{{{A_1} - {A_3}}}{{1 + 2{A_1} - {A_3}}}
\end{array} \right.
\eeq
can diagonalize the $H_{\mu\nu}$ into
\beq \label{eq:diagh}
H{'_{\mu \upsilon }} = \left( {\begin{array}{*{20}{c}}
{ - \frac{{1 + {A_1}}}{{1 + 2{A_1} - {A_3}}}}&{}&{}&{}\\
{}&1&{}&{}\\
{}&{}&1&{}\\
{}&{}&{}&{\frac{{1 + 2{A_1} - {A_3}}}{{1 + {A_1}}}}
\end{array}} \right)
\eeq
and transform the spurion into
\beq \label{eq:coodtrsmofn}
{N_\mu } = \left( {\begin{array}{*{20}{c}}
{\frac{1 + A_1}{1 + 2A_1 - A_3}}\\
{}\\
{}\\
1
\end{array}} \right).
\eeq
Therefore the $XDISIM2$ case can return to the case of $DISIM$ by a linear transformation in $t-z$ plane.

Meanwhile, the rank 3 conformal tensor is
\begin{widetext}
\beq \label{eq:r3cvtensor}
\begin{array}{l}
{{\tilde F}_{t\mu \upsilon }} = \left( {\begin{array}{*{20}{c}}
{3\left( {1 + {A_3}} \right)}&{}&{}&{1 + 3{A_3} - 2{A_1}}\\
{}&{ - \left( {1 + {A_1}} \right)}&{}&{}\\
{}&{}&{ - \left( {1 + {A_1}} \right)}&{}\\
{1 + 3{A_3} - 2{A_1}}&{}&{}&{ - 1 + 3{A_3} - 4{A_1}}
\end{array}} \right),\\
{{\tilde F}_{x\mu \upsilon }} = \left( {\begin{array}{*{20}{c}}
{}&{ - \left( {1 + {A_1}} \right)}&{}&{}\\
{ - \left( {1 + {A_1}} \right)}&{}&{}&{ - \left( {1 + {A_1}} \right)}\\
{}&{}&{}&{}\\
{}&{ - \left( {1 + {A_1}} \right)}&{}&{}
\end{array}} \right),
{{\tilde F}_{y\mu \upsilon }} = \left( {\begin{array}{*{20}{c}}
{}&{}&{ - \left( {1 + {A_1}} \right)}&{}\\
{}&{}&{}&{}\\
{ - \left( {1 + {A_1}} \right)}&{}&{}&{ - \left( {1 + {A_1}} \right)}\\
{}&{}&{ - \left( {1 + {A_1}} \right)}&{}
\end{array}} \right),\\
{{\tilde F}_{z\mu \upsilon }} = \left( {\begin{array}{*{20}{c}}
{1 + 3{A_3} - 2{A_1}}&{}&{}&{ - 1 + 3{A_3} - 4{A_1}}\\
{}&{ - \left( {1 + {A_1}} \right)}&{}&{}\\
{}&{}&{ - \left( {1 + {A_1}} \right)}&{}\\
{ - 1 + 3{A_3} - 4{A_1}}&{}&{}&{ - 3\left( {1 - {A_3} + 2{A_1}} \right)}
\end{array}} \right),
\end{array}
\eeq
\end{widetext}
with conformal factors ${e^{3{A_2}\theta }}$ and ${e^{2\left( {1+3A_3- 2{A_1}} \right)\theta }}$ under $R_z(\theta)$ and $B_z(\theta)$ respectively. However it is not an independent tensor for ${\tilde F_{\sigma \mu \upsilon }} = {N_{\left( \sigma  \right.}}{H_{\left. {\mu \upsilon } \right)}}$.

The existence of invariant metric demands no deformation of $r_z$, i.e. $A_2=0$. The invariant metric function is now
\beq \label{eq:invmtrxdism2}
{F^2} = {\left( {{N_\mu }{y^\mu }} \right)^{2\frac{{{A_1} - {A_3}}}{{1 + {A_1}}}}}{\left( {{H_{\mu \upsilon }}{y^\mu }{y^\upsilon }} \right)^{\frac{{1 + {A_3}}}{{1 + {A_1}}}}}.
\eeq

The deformed generators of $XDISIM2$ can be expressed with a free parameter,
\bw
\beq \label{eq:freeparexpxdisim2}
\begin{array}{l}
{r_z} = \left( {\begin{array}{*{20}{c}}
{{A_2}}&{}&{}&{}&{}\\
{}&{{A_2}}&{ - 1}&{}&{}\\
{}&1&{{A_2}}&{}&{}\\
{}&{}&{}&{{A_2}}&{}\\
{}&{}&{}&{}&0
\end{array}} \right),
{b_z} = \left( {\begin{array}{*{20}{c}}
{2\left( {\alpha  - {A_1}} \right)}&{}&{}&{1 + 2a - {A_3}}&{}\\
{}&{{A_3} - {A_1}}&{}&{}&{}\\
{}&{}&{{A_3} - {A_1}}&{}&{}\\
{1 + {A_3} + 2\left( {{A_1} - \alpha } \right)}&{}&{}&{2\left( {{A_3} - \alpha } \right)}&{}\\
{}&{}&{}&{}&0
\end{array}} \right),\\
{p_t} = \left( {\begin{array}{*{20}{c}}
0&{}&{}&{}&{1 + \alpha }\\
{}&0&{}&{}&{}\\
{}&{}&0&{}&{}\\
{}&{}&{}&0&{{A_1} - \alpha }\\
{}&{}&{}&{}&0
\end{array}} \right),
{p_z} = \left( {\begin{array}{*{20}{c}}
0&{}&{}&{}&{\alpha  - {A_1}}\\
{}&0&{}&{}&{}\\
{}&{}&0&{}&{}\\
{}&{}&{}&0&{1 + 2{A_1} - \alpha }\\
{}&{}&{}&{}&0
\end{array}} \right)\\
{p_x} = \left( {\begin{array}{*{20}{c}}
0&{}&{}&{}&{}\\
{}&0&{}&{}&{1 + {A_1}}\\
{}&{}&0&{}&{}\\
{}&{}&{}&0&{}\\
{}&{}&{}&{}&0
\end{array}} \right),
{p_y} = \left( {\begin{array}{*{20}{c}}
0&{}&{}&{}&{}\\
{}&0&{}&{}&{}\\
{}&{}&0&{}&{1 + {A_1}}\\
{}&{}&{}&0&{}\\
{}&{}&{}&{}&0
\end{array}} \right).
\end{array}
\eeq
\ew
Hence, the conformal covariant rank 1 tensor is spurion $N_{\mu}$ with the conformal factor as in the previous representation. The rank 2 conformal covariant tensor is
\beq \label{eq:r2tensorfreexdisim2}
{\tilde H_{\mu \upsilon }} = \left( {\begin{array}{*{20}{c}}
{ - \frac{{1 + {A_3} + 2{A_1} - 2\alpha }}{{1 + {A_1}}}}&{}&{}&{\frac{{2\alpha  - {A_1} - {A_3}}}{{1 + {A_1}}}}\\
{}&1&{}&{}\\
{}&{}&1&{}\\
{\frac{{2\alpha  - {A_1} - {A_3}}}{{1 + {A_1}}}}&{}&{}&{\frac{{1 + 2\alpha  - {A_3}}}{{1 + {A_1}}}}
\end{array}} \right)
\eeq
with conformal factor the same as \eqref{eq:hforxdisim2}.

\subsubsection{$ISIM$ group}

The rank 1 conformal covariant tensor $N_{\mu}$ for undeformed $SIM$ group has conformal factor $e^{\theta}$ under the action of $B_z$, while the rank 2 conformal covariant tensor $G_{\mu\nu}$ is invariant under $B_z$. The fully symmetric rank 3 conformal covariant tensor is
\beq \label{eq:r3forisim}
\begin{array}{l}
{T_{1\mu \upsilon }} = \left( {\begin{array}{*{20}{c}}
{3a}&{}&{}&a\\
{}&{ - a}&{}&{}\\
{}&{}&{ - a}&{}\\
a&{}&{}&{ - a}
\end{array}} \right),
{T_{2\mu \upsilon }} = \left( {\begin{array}{*{20}{c}}
{}&{ - a}&{}&{}\\
{ - a}&{}&{}&{ - a}\\
{}&{}&{}&{}\\
{}&{ - a}&{}&{}
\end{array}} \right),\\
{T_{4\mu \upsilon }} = \left( {\begin{array}{*{20}{c}}
a&{}&{}&{ - a}\\
{}&{ - a}&{}&{}\\
{}&{}&{ - a}&{}\\
{ - a}&{}&{}&{ - 3a}
\end{array}} \right),
{T_{3\mu \upsilon }} = \left( {\begin{array}{*{20}{c}}
{}&{}&{ - a}&{}\\
{}&{}&{}&{}\\
{ - a}&{}&{}&{ - a}\\
{}&{}&{ - a}&{}
\end{array}} \right).
\end{array}
\eeq
However the rank 3 tensor is not independent, it can be decomposed into the direct product of rank 1 and rank 2 tensors. The metric function is therefore of the form
\beq \label{eq:mtrformisim}
{F^2} = {\left( {{N_\mu }{y^\mu }} \right)^A}{\left( {{G_{\mu \upsilon }}{y^\mu }{y^\upsilon }} \right)^B}.
\eeq
The invariance of $F^2$ demands $A=0$. Finally the metric function is
\beq \label{eq:smplmtrformisim}
{F^2} =  G_{\mu \upsilon }{y^\mu }{y^\upsilon} .
\eeq

\subsection{The $DIHOM$ group}

There are two subclasses for $DIHOM$, one subclass denoted by $WDISM$ which has the same corresponding structure with $DISIM$, while the other denoted by $DIHOM$ which is totally different from $DISIM$.

For the case of $XDISIM$ which is lack of $r_z$, the deformed generator is only $b_z$. The result is the same as $DISIMb$. Note that the rank 2 conformal covariant tensor has an additional form,
\beq \label{eq:addr2tenforxdisim}
{F_{\mu \upsilon }} = \left( {\begin{array}{*{20}{c}}
{}&a&b&{}\\
{ - a}&{}&{}&{ - a}\\
{ - b}&{}&{}&{ - b}\\
{}&a&b&{}
\end{array}} \right),
\eeq
which is the only difference needed to be noticed. The tensor is antisymmetric so it is not appropriate to construct invariant metric.

The deformed generators of $DIHOM$ are
\beq \label{eq:dfmgendihom}
\begin{array}{l}
{t_2} = \left( {\begin{array}{*{20}{c}}
{}&{}&1&{}&{}\\
{}&{}&{}&{}&{}\\
1&{}&{}&1&{}\\
{}&{}&{ - 1}&{}&{}\\
{ - \left( {{A_1} + {A_2}} \right)}&{}&{}&{ - \left( {{A_1} + {A_2}} \right)}&0
\end{array}} \right),\\
{p_t} = \left( {\begin{array}{*{20}{c}}
0&{}&{}&{}&1\\
{}&0&{}&{}&{}\\
{ - \left( {{A_1} + {A_2}} \right)}&{}&0&{}&{}\\
{}&{}&{}&0&{}\\
{}&{}&{}&{}&0
\end{array}} \right),\\
{p_x} = \left( {\begin{array}{*{20}{c}}
0&{}&{}&{}&{}\\
{}&0&{}&{}&1\\
{}&{{A_1} + {A_2}}&0&{}&{}\\
{}&{}&{}&0&{}\\
{}&{}&{}&{}&0
\end{array}} \right),\\
{p_y} = \left( {\begin{array}{*{20}{c}}
0&{}&{}&{{A_1}}&{}\\
{}&0&{}&{}&{}\\
{}&{}&{{A_1} + {A_2}}&{}&1\\
{{A_1}}&{}&{}&0&{}\\
{}&{}&{}&{}&{ - \left( {{A_1} + {A_2}} \right)}
\end{array}} \right),\\
{p_z} = \left( {\begin{array}{*{20}{c}}
0&{}&{}&{}&{}\\
{}&0&{}&{}&{}\\
{}&{}&0&{{A_1} + {A_2}}&{}\\
{}&{}&{}&0&1\\
{}&{}&{}&{}&0
\end{array}} \right).
\end{array}
\eeq
$N_{\mu}$ is still the rank 1 conformal covariant tensor with conformal factor $e^{\theta}$ and $e^{A_1\theta}$ under the action of $B_z$ and $p_y$. However, except ${N_\mu }{N_\upsilon }$ the rank 2 conformal covariant tensor for $DIHOM$ is only
\beq \label{eq:r2tendihom}
{\tilde F_{\mu \upsilon }} = \left( {\begin{array}{*{20}{c}}
{}&1&{}&{}\\
{ - 1}&{}&{}&{ - 1}\\
{}&{}&{}&{}\\
{}&1&{}&{}
\end{array}} \right),
\eeq
which is antisymmetric and not appropriate to construct invariant metric. Therefore the invariant metric for $DIHOM$ does not exist.

For the undeformed $HOM$ group, the rank 3 conformal covariant tensor ${T_{\sigma \mu \upsilon }}$ under $SIM$ is also conformal covariant under $HOM$. So \eqref{eq:mtrformisim} is also the invariant tensor for $IHOM$.

\subsection{$DTE$ group}

There are three classes, where the second and third classes can be specified into two subclasses.
\subsubsection{$DTE1$ group}

The deformed generator relative to the semidirect of $E(2)$ and $T(4)$ is only $r_z$,
\beq \label{eq:dfmgendte1}
{r_z} = \left( {\begin{array}{*{20}{c}}
{{A_1}}&{}&{}&{ - {A_2}}&{}\\
{}&{{A_1} + {A_2}}&{ - 1}&{}&{}\\
{}&1&{{A_1} + {A_2}}&{}&{}\\
{{A_2}}&{}&{}&{{A_1} + 2{A_2}}&{}\\
{}&{}&{}&{}&0
\end{array}} \right).
\eeq
The corresponding single parameter group element is
\beq \label{eq:dfmgrpeledte1}
{R_z}\left( \theta  \right) = {e^{\left( {{A_1} + {A_2}} \right)\theta }}\left( {\begin{array}{*{20}{c}}
{1 - {A_2}\theta }&{}&{}&{ - {A_2}\theta }\\
{}&{\cos \theta }&{ - \sin \theta }&{}\\
{}&{\sin \theta }&{\cos \theta }&{}\\
{{A_2}\theta }&{}&{}&{1 + {A_2}\theta }
\end{array}} \right).
\eeq
The spurion $N_{\mu}$ is still the rank 1 conformal covariant tensor of $DTE1$ with conformal factor ${e^{\left( {{A_1} + {A_2}} \right)\theta }}$ under the action of $R_z$. The rank 2 conformal covariant tensor of $DTE1$ is only ${N_\mu }{N_\upsilon }$. Hence the invariant metric under the action of $DTE1$ does not exist.

However, in the case of ${A_2} = 0$, there exists a kind of rank 2 conformal covariant tensor,
\beq \label{eq:r2tensordte1}
{G_{\mu \upsilon }} = \left( {\begin{array}{*{20}{c}}
a&{}&{}&b\\
{}&{b - a}&{}&{}\\
{}&{}&{b - a}&{}\\
b&{}&{}&{2b - a}
\end{array}} \right),
\eeq
with the conformal factor ${e^{2{A_1}\theta }}$ under the action of $R_z$. As in  \eqref{eq:cstronmtr} the first and second equation will force $A_1=0$ and hence there does not exist $DTE1$ invariant metric in all the cases. However there can exist the Minkowski-Finsler type of geometry under the action of $IE(2)$.

\subsubsection{$DTE2$ group}

$DTE2$ has two subclasses $DTE2a$ and $DTE2b$. $DTE2b$ does not have the representation inherited from the 5-d representation of Poincar\'e group. $DTE2a$ has two different kind of natural representations.

In the first natural representation of $DTE2a$, the matrices of four deformed translation generators are
\beq \label{eq:mtrofpdte2a}
\begin{array}{c}
{p_t} = \left( {\begin{array}{*{20}{c}}
{2\beta  - \frac{{{A_2}}}{2}}&{}&{}&{ - \frac{{{A_2}}}{2}}&1\\
{}&\beta &{}&{}&{}\\
{}&{}&\beta &{}&{}\\
{ - \frac{{{A_2}}}{2}}&{}&{}&{2\beta  - \frac{{{A_2}}}{2}}&{}\\
{}&{}&{}&{}&0
\end{array}} \right),\\
{p_x} = \left( {\begin{array}{*{20}{c}}
{}&{ - \beta }&{}&{}&{}\\
{\beta  - {A_2}}&{}&{}&{\beta  - {A_2}}&1\\
{}&{}&{}&{}&{}\\
{}&\beta &{}&{}&{}\\
{}&{}&{}&{}&0
\end{array}} \right),\\
{p_z} = \left( {\begin{array}{*{20}{c}}
{\frac{{{A_2}}}{2}}&{}&{}&{\frac{{{A_2}}}{2} - 2\beta }&{}\\
{}&\beta &{}&{}&{}\\
{}&{}&\beta &{}&{}\\
{2\beta  - \frac{{3{A_2}}}{2}}&{}&{}&{4\beta  - \frac{{3{A_2}}}{2}}&1\\
{}&{}&{}&{}&0
\end{array}} \right),\\
{p_y} = \left( {\begin{array}{*{20}{c}}
{}&{}&{ - \beta }&{}&{}\\
{}&{}&{}&{}&{}\\
{\beta  - {A_2}}&{}&{}&{\beta  - {A_2}}&1\\
{}&{}&\beta &{}&{}\\
{}&{}&{}&{}&0
\end{array}} \right),
\end{array}
\eeq
where $\beta$ satisfies ${\beta ^2} - {A_2}\beta  + {A_1} = 0$.

The spurion $N_{\mu}$ is still the rank 1 conformal covariant tensor here with conformal factor ${e^{\left( {2\beta  - {A_2}} \right)\theta }}$ and ${e^{\left( {2\beta  - {A_2}} \right)\theta }}$ under the action of $P_t$ and $P_z$ respectively, while the independent rank 2 conformal covariant tensor rather than ${N_\mu }{N_\upsilon }$ only exists in the case ${A_2} = 2\beta $  and is the Minkowski metric tensor ${G_{\mu \upsilon }}$ which has conformal factor ${e^{\left( {4\beta  - {A_2}} \right)\theta }}$ and ${e^{2\beta \theta }}$ under the action of $P_t$ and $P_z$ respectively. In the case ${A_2} = 2\beta $, $N_{\mu}$ is invariant while ${G_{\mu \upsilon }}$ is scaled by factor ${e^{2\beta \theta }}$ under the group action.

The invariant metric under $DTE2a$ in the present representation can only be of the form
\beq \label{eq:mtrinvdte2a1}
F = {N_\mu }{y^\mu }.
\eeq

In the second natural representation of $DTE2a$, also only translation generators deform with the natural representation matrices are
\beq \label{eq:mtrofpdte2a2}
\begin{array}{l}
{p_t} = \left( {\begin{array}{*{20}{c}}
\lambda &{}&{}&{\lambda  - {A_2}}&1\\
{}&\lambda &{}&{}&{}\\
{}&{}&\lambda &{}&{}\\
{\lambda  - {A_2}}&{}&{}&\lambda &{}\\
{}&{}&{}&{}&0
\end{array}} \right),\\
{p_x} = \left( {\begin{array}{*{20}{c}}
{}&{\lambda  - {A_2}}&{}&{}&{}\\
{\lambda  - {A_2}}&{}&{}&{\lambda  - {A_2}}&1\\
{}&{}&{}&{}&{}\\
{}&{{A_2} - \lambda }&{}&{}&{}\\
{}&{}&{}&{}&0
\end{array}} \right),\\
{p_z} = \left( {\begin{array}{*{20}{c}}
\lambda &{}&{}&{\lambda  - {A_2}}&{}\\
{}&\lambda &{}&{}&{}\\
{}&{}&\lambda &{}&{}\\
{\lambda  - {A_2}}&{}&{}&\lambda &{}\\
{}&{}&{}&{}&0
\end{array}} \right),\\
{p_y} = \left( {\begin{array}{*{20}{c}}
{}&{}&{\lambda  - {A_2}}&{}&{}\\
{}&{}&{}&{}&{}\\
{\lambda  - {A_2}}&{}&{}&{\lambda  - {A_2}}&1\\
{}&{}&{{A_2} - \lambda }&{}&{}\\
{}&{}&{}&{}&0
\end{array}} \right),
\end{array}
\eeq
where $\lambda$ satisfies ${\lambda ^2} - {A_2}\lambda  + {A_1} = 0$.

In the present presentation, the spurion $N_{\mu}$ is still the rank 1 conformal covariant tensor with conformal factor ${e^{\left( {2\lambda  - {A_2}} \right)\theta }}$ under the action of $P_t$ and $P_z$. The form of rank 2 conformal covariant tensor under the action $DTE2a$ depends on the $\lambda$ value. When $\lambda  = {A_2}$, the rank 2 tensor is
\beq \label{eq:r21dte2a22}
{H_{\mu \upsilon }} = \left( {\begin{array}{*{20}{c}}
a&{}&{}&{a + b}\\
{}&b&{}&{}\\
{}&{}&b&{}\\
{a + b}&{}&{}&{a + 2b}
\end{array}} \right),
\eeq
while it can only be Minkowski metric tensor ${G_{\mu \upsilon }}$ when $\lambda  \ne {A_2}$. The conformal factor is ${e^{2\lambda \theta }}$ in both cases.

Concerning the construction of the $DTE2a$ invariant metric function, in the case $\lambda  \ne {A_2}$, the invariant metric function has the form
\beq \label{eq:mtrinvdte2a22}
{F^2} = {\left( {{G_{\mu \upsilon }}{y^\mu }{y^\upsilon }} \right)^a}{\left( {{N_\mu }{y^\mu }} \right)^b},
\eeq
with the constrain equation
\beq \label{eq:cstrmtrinvdte2a22}
\left\{ \begin{array}{l}
2a + b = 2\\
2a\lambda  + b\left( {2\lambda  - {A_2}} \right) = 0
\end{array} \right.,
\eeq
and solution
\beq \label{eq:solnmtrinvdte2a22}
\left\{ \begin{array}{l}
a = \frac{{{A_2} - 2\lambda }}{{{A_2} - \lambda }}\\
b = \frac{{2\lambda }}{{{A_2} - \lambda }}
\end{array} \right..
\eeq
The metric function is therefore of the form finally
\beq \label{eq:metricinvdte2a22}
{F^2} = {\left( {{G_{\mu \upsilon }}{y^\mu }{y^\upsilon }} \right)^{\frac{{{A_2} - 2\lambda }}{{{A_2} - \lambda }}}}{\left( {{N_\mu }{y^\mu }} \right)^{\frac{{2\lambda }}{{{A_2} - \lambda }}}}.
\eeq
There does not exist $DTE2a$ invariant metric function in the case $\lambda  = {A_2}$.

\subsubsection{$DTE3$ group}

Like $DTE2$, the $DTE3$ can also be specified into two subclasses, $DTE3a$ and $DTE3b$. The representation of $DTE3a$ is the same as $DTE2a$ and hence so is the invariant metric function.

The natural representation of deformed generators in $DTE3b$ is
\beq \label{eq:mtrrepdte3b}
\begin{array}{c}
{p_t} = \left( {\begin{array}{*{20}{c}}
{}&{}&{}&{{A_1}}&1\\
{}&{ - {A_1}}&{}&{}&{}\\
{}&{}&{ - {A_1}}&{}&{}\\
{ - {A_1}}&{}&{}&{ - 2{A_1}}&{}\\
{}&{}&{}&{}&0
\end{array}} \right),\\
{r_z} = \left( {\begin{array}{*{20}{c}}
{{A_2}}&{}&{}&{{A_2}}&{}\\
{}&{}&{ - 1}&{}&{}\\
{}&1&{}&{}&{}\\
{ - {A_2}}&{}&{}&{ - {A_2}}&{}\\
{}&{}&{}&{}&0
\end{array}} \right),\\
{p_z} = \left( {\begin{array}{*{20}{c}}
{}&{}&{}&{{A_1}}&{}\\
{}&{ - {A_1}}&{}&{}&{}\\
{}&{}&{ - {A_1}}&{}&{}\\
{ - {A_1}}&{}&{}&{ - 2{A_1}}&1\\
{}&{}&{}&{}&0
\end{array}} \right),\\
{p_x} = \left( {\begin{array}{*{20}{c}}
{}&{ - {A_1}}&{}&{}&{}\\
{ - {A_1}}&{}&{}&{ - {A_1}}&1\\
{}&{}&{}&{}&{}\\
{}&{{A_1}}&{}&{}&{}\\
{}&{}&{}&{}&0
\end{array}} \right).
\end{array}
\eeq
The rank 1 conformal covariant tensor is the spurion $N_{\mu}$ with conformal factor ${e^{ - {A_1}\theta }}$ under the action of $P_t$ and $P_z$. The rank 2 conformal covariant tensor can only be ${N_\mu }{N_\upsilon }$ in the case of ${A_1} \ne 0$. When ${A_1} = 0$ it can have different form
\beq \label{eq:r2formdte3b}
{H_{\mu \upsilon }} = \left( {\begin{array}{*{20}{c}}
a&{}&{}&b\\
{}&{b - a}&{}&{}\\
{}&{}&{b - a}&{}\\
b&{}&{}&{2b - a}
\end{array}} \right),
\eeq
which is invariant under $DTE3b$.

The construction of the invariant metric function has thus plenty of variety here. The phenomena is similar to $IE(2)$ and will be discussed in the invariant metric function of $IE(2)$.

\subsubsection{$IE(2)$ group}

$IE(2)$ group is specific in the aspect that it does not admit only Riemannian structure but also Finslerian structure of spacetime. The rank 2 invariant tensor under $IE(2)$ is of the form
\beq \label{eq:r2formie2}
{G_{\left( {a,b} \right)\mu \upsilon }} = \left( {\begin{array}{*{20}{c}}
a&{}&{}&{a + b}\\
{}&b&{}&{}\\
{}&{}&b&{}\\
{a + b}&{}&{}&{a + 2b}
\end{array}} \right),
\eeq
where $a$ and $b$ are free parameters. In case of $b=0$, it reduces to ${N_\mu }{N_\upsilon }$ while it gives the Minkowski metric tensor in case of $b=-a$. Because of ${G_{\left( {a,b} \right)\mu \upsilon }}$ is an invariant rank 2 tensor, the construction of invariant metric function is thus a little bit of arbitrary in the sense that
\beq \label{eq:mtrformie2}
{F^2} = \prod\limits_{a,b} {{{\left( {{G_{\left( {a,b} \right)\mu \upsilon }}{y^\mu }{y^\upsilon }} \right)}^{{D_{a,b}}}}},
\eeq
where it is only need to satisfy the constrain condition
\beq \label{eq:cstrmtrie2}
\sum\limits_{a,b} {{D_{a,b}}}  = 1,
\eeq
e.g.
\beq \label{eq:expmtrie2}
{F^2} = \frac{{{{\left( {{G_{\left( { - 1,1} \right)\mu \upsilon }}{y^\mu }{y^\upsilon }} \right)}^2}}}{{{G_{\left( {1,1} \right)\mu \upsilon }}{y^\mu }{y^\upsilon }}}
\eeq
is an admissible form of invariant metric function.

The only $IE(2)$ invariant Riemannian metric function is ${F^2} = {G_{\left( {a,b} \right)\mu \upsilon }}{y^\mu }{y^\upsilon }$ while there are many in Finslerian type.

\subsection{$DISO(3)$ group}

There are two classes in classification of deformed group $DISO(3)$ of $SO(3)$, $DISO(3)1$ which has only one natural representation and $DISO(3)2$ which has three inequivalent natural representations.

\subsubsection{$DISO(3)1$ group}

The matrices of deformed generators of $DISO(3)1$ are
\beq \label{eq:mtrrepdiso31}
\begin{array}{c}
{p_t} = \left( {\begin{array}{*{20}{c}}
\alpha &{}&{}&{}&1\\
{}&\alpha &{}&{}&{}\\
{}&{}&\alpha &{}&{}\\
{}&{}&{}&\alpha &{}\\
{}&{}&{}&{}&0
\end{array}} \right),
{p_x} = \left( {\begin{array}{*{20}{c}}
{}&\beta &{}&{}&{}\\
\alpha &{}&{}&{}&1\\
{}&{}&0&{}&{}\\
{}&{}&{}&0&{}\\
{}&{}&{}&{}&0
\end{array}} \right),\\
{p_y} = \left( {\begin{array}{*{20}{c}}
{}&{}&\beta &{}&{}\\
{}&0&{}&{}&{}\\
\alpha &{}&{}&{}&1\\
{}&{}&{}&0&{}\\
{}&{}&{}&{}&0
\end{array}} \right),
{p_z} = \left( {\begin{array}{*{20}{c}}
{}&{}&{}&\beta &{}\\
{}&0&{}&{}&{}\\
{}&{}&0&{}&{}\\
a&{}&{}&{}&1\\
{}&{}&{}&{}&0
\end{array}} \right),
\end{array}
\eeq
where $\alpha$ and $\beta$ satisfy $\alpha \beta  + {A_1} = 0$.

There does not exist any rank 1 conformal covariant tensor under the action of $DISO(3)1$. The rank 2 conformal covariant tensor is only Minkowski metric tensor with the conformal factor ${e^{2\alpha \theta }}$ under the action of ${P_t}$. There does not exist $DISO(3)1$ invariant metric function.

\subsubsection{$DISO(3)2$ group}

There are three inequivalent natural representation of $DISO(3)2$ group.

In the first representation, only $p_t$ is deformed
\beq \label{eq:mtrrepdiso32}
{p_t} = \left( {\begin{array}{*{20}{c}}
{{A_1}}&{}&{}&{}&1\\
{}&{{A_1}}&{}&{}&{}\\
{}&{}&{{A_1}}&{}&{}\\
{}&{}&{}&{{A_1}}&{}\\
{}&{}&{}&{}&0
\end{array}} \right).
\eeq
The rank 1 conformal covariant tensor is not the spurion $N_{\mu}$ anymore but
\beq \label{eq:r1diso32}
{M_\mu } = \left( {\begin{array}{*{20}{c}}
1\\
0\\
0\\
0
\end{array}} \right)
\eeq
with conformal factor ${e^{ {A_1}\theta }}$ under the action of $P_t$.
The rank 2 conformal covariant tensor is
\beq \label{eq:r2diso32}
{G_{\mu \upsilon }} = \left( {\begin{array}{*{20}{c}}
a&{}&{}&{}\\
{}&b&{}&{}\\
{}&{}&b&{}\\
{}&{}&{}&b
\end{array}} \right)
\eeq
with conformal factor ${e^{ 2{A_1}\theta }}$ under the action of $P_t$. There does not exist invariant metric function under group action.

In the second representation, matrices for deformed generators are
\beq \label{eq:mtrrepdiso322}
\begin{array}{c}
{p_t} = \left( {\begin{array}{*{20}{c}}
{ - {A_1}}&{}&{}&{}&1\\
{}&0&{}&{}&{}\\
{}&{}&0&{}&{}\\
{}&{}&{}&0&{}\\
{}&{}&{}&{}&0
\end{array}} \right),
{p_x} = \left( {\begin{array}{*{20}{c}}
0&{}&{}&{}&{}\\
{ - {A_1}}&0&{}&{}&1\\
{}&{}&0&{}&{}\\
{}&{}&{}&0&{}\\
{}&{}&{}&{}&0
\end{array}} \right),\\
{p_y} = \left( {\begin{array}{*{20}{c}}
0&{}&{}&{}&{}\\
{}&0&{}&{}&{}\\
{ - {A_1}}&{}&0&{}&1\\
{}&{}&{}&0&{}\\
{}&{}&{}&{}&0
\end{array}} \right),
{p_x} = \left( {\begin{array}{*{20}{c}}
0&{}&{}&{}&{}\\
{}&0&{}&{}&{}\\
{}&{}&0&{}&{}\\
{ - {A_1}}&{}&{}&0&1\\
{}&{}&{}&{}&0
\end{array}} \right).
\end{array}
\eeq
The rank 1 conformal covariant tensor is $M_{\mu}$ in \eqref{eq:r1diso32} with conformal factor ${e^{ -{A_1}\theta }}$ under the action of $P_t$. There does not exist any rank 2 conformal covariant tensor. Neither does there the $DISO(3)2$ group invariant metric function.

The same happens to the third representation of $DISO(3)2$ group.

\subsubsection{$ISO(3)$ group}

The $ISO(3)$ invariant rank 1 tensor is $M_{\mu}$ and rank 2 tensor is
\beq \label{eq:r2iso3}
{G_{\left( {a,b} \right)\mu \upsilon }} = \left( {\begin{array}{*{20}{c}}
a&{}&{}&{}\\
{}&b&{}&{}\\
{}&{}&b&{}\\
{}&{}&{}&b
\end{array}} \right).
\eeq
The invariant metric function is hence
\beq \label{eq:invmtriso3}
{F^2} = {\left( {{M_\mu }{y^\mu }} \right)^A}\prod\limits_{a,b} {{{\left( {{G_{\left( {a,b} \right)\mu \upsilon }}{y^\mu }{y^\upsilon }} \right)}^{{B_{a,b}}}}},
\eeq
with the corresponding constrain condition
\beq \label{eq:cstrinvmtriso3}
A + 2\sum\limits_{a,b} {{B_{a,b}}}  = 2.
\eeq

\subsection{$DISO(2,1)$ group}

There are two classes in the classification of $DISO(2,1)$, $DISO(2,1)1$ which has only one natural representation and $DISO(2,1)2$ which has two inequivalent natural representations.

\subsubsection{$DISO(2,1)1$ group}

The matrices for deformed generators of $DISO(2,1)1$ are
\beq \label{eq:matrepdiso211}
\begin{array}{l}
{p_x} = \left( {\begin{array}{*{20}{c}}
\alpha &{}&{}&{}&{}\\
{}&\alpha &{}&{}&1\\
{}&{}&\alpha &{}&{}\\
{}&{}&{}&\alpha &{}\\
{}&{}&{}&{}&0
\end{array}} \right),
{p_t} = \left( {\begin{array}{*{20}{c}}
{}&\alpha &{}&{}&1\\
\beta &{}&{}&{}&{}\\
{}&{}&0&{}&{}\\
{}&{}&{}&0&{}\\
{}&{}&{}&{}&0
\end{array}} \right),\\
{p_y} = \left( {\begin{array}{*{20}{c}}
0&{}&{}&{}&{}\\
{}&{}&{ - \beta }&{}&{}\\
{}&\alpha &{}&{}&1\\
{}&{}&{}&0&{}\\
{}&{}&{}&{}&0
\end{array}} \right),
{p_z} = \left( {\begin{array}{*{20}{c}}
0&{}&{}&{}&{}\\
{}&{}&{}&{ - \beta }&{}\\
{}&{}&0&{}&{}\\
{}&\alpha &{}&{}&1\\
{}&{}&{}&{}&0
\end{array}} \right),
\end{array}
\eeq
where $\alpha$ and $\beta$ satisfy $\alpha \beta  + {A_1} = 0$. Like in $DISO(3)1$, we can specify two cases to discuss.

When $A_1>0$, we can take $\beta =-\alpha$ and get
\beq \label{eq:rep1diso211}
\begin{array}{l}
{p_x} = \left( {\begin{array}{*{20}{c}}
\alpha &{}&{}&{}&{}\\
{}&\alpha &{}&{}&1\\
{}&{}&\alpha &{}&{}\\
{}&{}&{}&\alpha &{}\\
{}&{}&{}&{}&0
\end{array}} \right),
{p_t} = \left( {\begin{array}{*{20}{c}}
{}&\alpha &{}&{}&1\\
-\alpha &{}&{}&{}&{}\\
{}&{}&0&{}&{}\\
{}&{}&{}&0&{}\\
{}&{}&{}&{}&0
\end{array}} \right),\\
{p_y} = \left( {\begin{array}{*{20}{c}}
0&{}&{}&{}&{}\\
{}&{}&{ \alpha }&{}&{}\\
{}&\alpha &{}&{}&1\\
{}&{}&{}&0&{}\\
{}&{}&{}&{}&0
\end{array}} \right),
{p_z} = \left( {\begin{array}{*{20}{c}}
0&{}&{}&{}&{}\\
{}&{}&{}&{ \alpha }&{}\\
{}&{}&0&{}&{}\\
{}&\alpha &{}&{}&1\\
{}&{}&{}&{}&0
\end{array}} \right),
\end{array}
\eeq
where $\alpha  =  \pm \sqrt {{A_1}}$.

There does not exist rank 1 conformal covariant tensor in this case. The conformal covariant rank 2 tensor is just the Minkowski metric tensor.

When $A_1<0$, we can take $\beta =\alpha$ and get
\beq \label{eq:rep2diso211}
\begin{array}{l}
{p_x} = \left( {\begin{array}{*{20}{c}}
\alpha &{}&{}&{}&{}\\
{}&\alpha &{}&{}&1\\
{}&{}&\alpha &{}&{}\\
{}&{}&{}&\alpha &{}\\
{}&{}&{}&{}&0
\end{array}} \right),
{p_t} = \left( {\begin{array}{*{20}{c}}
{}&\alpha &{}&{}&1\\
\alpha &{}&{}&{}&{}\\
{}&{}&0&{}&{}\\
{}&{}&{}&0&{}\\
{}&{}&{}&{}&0
\end{array}} \right),\\
{p_y} = \left( {\begin{array}{*{20}{c}}
0&{}&{}&{}&{}\\
{}&{}&{ -\alpha }&{}&{}\\
{}&\alpha &{}&{}&1\\
{}&{}&{}&0&{}\\
{}&{}&{}&{}&0
\end{array}} \right),
{p_z} = \left( {\begin{array}{*{20}{c}}
0&{}&{}&{}&{}\\
{}&{}&{}&{ -\alpha }&{}\\
{}&{}&0&{}&{}\\
{}&\alpha &{}&{}&1\\
{}&{}&{}&{}&0
\end{array}} \right),
\end{array}
\eeq
where $\alpha  =  \pm \sqrt {{-A_1}}$.
There does not exist rank 1 conformal covariant tensor. The conformal covariant rank 2 tensor is
\beq \label{eq:r2covtensordiso211}
{H_{\mu \upsilon }} = \left( {\begin{array}{*{20}{c}}
{ - 1}&{}&{}&{}\\
{}&{ - 1}&{}&{}\\
{}&{}&1&{}\\
{}&{}&{}&1
\end{array}} \right).
\eeq

In both of two cases, the rank 2 tensor have the conformal factor ${e^{2\alpha \theta }}$ under ${P_x}$ and therefore are not appropriate to construct the invariant metric function.
\subsubsection{$DISO(2,1)2$ group}

$DISO(2,1)2$ group has two inequivalent natural representations.

In the first representation, the rank 1 conformal covariant tensor is
\beq \label{eq:r1covtensordiso212}
{M_\mu } = \left( {\begin{array}{*{20}{c}}
0\\
1\\
0\\
0
\end{array}} \right),
\eeq
with the conformal factor ${e^{ - {A_1}\theta }}$ under $P_x$ while the rank 2 conformal covariant tensor is
\beq \label{eq:r2covtensordiso212}
{H_{\mu \upsilon }} = \left( {\begin{array}{*{20}{c}}
a&{}&{}&{}\\
{}&b&{}&{}\\
{}&{}&{ - a}&{}\\
{}&{}&{}&{ - a}
\end{array}} \right),
\eeq
with the conformal factor ${e^{ - 2{A_1}\theta }}$ under $P_x$. So this representation does not give invariant metric function.

In the second representation, the rank 1 conformal covariant tensor is still $M_\mu $ of \eqref{eq:r1covtensordiso212} with the conformal factor ${e^{  {A_1}\theta }}$ while there does not exist the rank 2 conformal covariant tensor. So there still does not exist invariant metric function in this representation.

\subsubsection{$ISO(2,1)$ group}

The rank 1 conformal invariant tensor under $ISO(2,1)$ group action is $M_\mu $ of \eqref{eq:r1covtensordiso212} while the rank 2 invariant tensor is
\beq \label{eq:r2covtensoriso21}
{H_{\left( {a,b} \right)\mu \upsilon }} = \left( {\begin{array}{*{20}{c}}
a&{}&{}&{}\\
{}&b&{}&{}\\
{}&{}&{ - a}&{}\\
{}&{}&{}&{ - a}
\end{array}} \right).
\eeq
The group action invariant metric function is therefore
\beq \label{eq:invmtriso21}
{F^2} = {\left( {{M_\mu }{y^\mu }} \right)^A}\prod\limits_{a,b} {{{\left( {{H_{\left( {a,b} \right)\mu \upsilon }}{y^\mu }{y^\upsilon }} \right)}^{{B_{a,b}}}}},
\eeq
with the constrain condition
\beq \label{eq:cstriso21}
A + 2\sum\limits_{a,b} {{B_{a,b}}}  = 2.
\eeq

\subsection{Summary}

We summarize the metric functions with respect to various of Poincar\'e subgroups and the deformed Poincar\'e subgroups in Table I.

%\begin{widetext}
\begin{table}

\begin{center}
\begin{tabular}{c|c|c}
\end{tabular}
\end{center}
\end{table}
%\end{widetext}

%\begin{widetext}
\begin{table}
\begin{center}

\caption{The Finsler spacetime metric functions with symmetry group of semi-product Poincar\'e subgroups and their deformed partners}
\begin{tabular}{c|c|c}

\hline
\hline
\multirow{2}{*}{symmetry} & conformal covariant & conformal\\
 & tensor & factor\\
\cline{2-3}
\multirow{2}{*}{group} & \multicolumn{2}{|c}{the invariant metric and}\\
 & \multicolumn{2}{|c}{additional remark}\\
  \hline
  \hline
\multirow{2}{*}{de Sitter} & \multicolumn{2}{|c}{no conformal covariant tensor }\\
 &  \multicolumn{2}{|c}{and conformal factor}\\
\hline

\multirow{7}{*}{Poincar\'e} & \multirow{5}{*}{$G_{\mu \upsilon } = \left( {\begin{array}{*{20}{c}}
{ - 1}&{}&{}&{}\\
{}&1&{}&{}\\
{}&{}&1&{}\\
{}&{}&{}&1
\end{array}} \right)$} & \multirow{5}{*}{invariant}\\
 & & \\
 & & \\
 & & \\
 & & \\
\cline{2-3}
 & \multicolumn{2}{|c}{\multirow{2}{*}{${F^2} = {G_{\mu \upsilon }}{y^\mu }{y^\upsilon }$}}\\
 & \multicolumn{2}{|c}{}\\
 \hline
 \multirow{10}{*}{$DISIM$} & \multirow{4}{*}{${N_\mu } = {\left( {\begin{array}{*{20}{c}}
1&0&0&1
\end{array}} \right)^{\bf{T}}}$} & \multirow{2}{*}{${B_z}\left( \theta  \right):$}\\
 & & \\
 & &\multirow{2}{*}{ ${e^{\left( {1 + {A_2}} \right)\theta }}$}\\
  & & \\
\cline{2-3}
 & \multirow{4}{*}{${G_{\mu \upsilon }}$} & \multirow{2}{*}{${B_z}\left( \theta  \right):$} \\
 & & \\
  & & \multirow{2}{*}{${e^{ 2{A_1} \theta }}$}\\
   & & \\
\cline{2-3}
 & \multicolumn{2}{|c}{\multirow{2}{*}{${F^2} = {\left( {{G_{\mu \upsilon }}{y^\mu }{y^\upsilon }} \right)^{1 + {A_2}}}{\left( {{N_\mu }{y^\mu }} \right)^{ - 2{A_2}}}$}}\\
  & \multicolumn{2}{|c}{}\\
\hline
\multirow{12}{*}{$XDISIM1$} & \multirow{4}{*}{${N_\mu } $} & \multirow{2}{*}{${B_z}\left( \theta  \right):$}\\
& & \\
& &\multirow{2}{*}{${e^{\left( {1 + {A_3}} \right)\theta }}$} \\
& & \\
\cline{2-3}
 & \multirow{4}{*}{${G_{\mu \upsilon }}$} & \multirow{2}{*}{${B_z}\left( \theta  \right):$} \\
 & & \\
& & \multirow{2}{*}{${e^{2\left( {{A_3} - {A_1}} \right)\theta }}$}\\
& & \\
\cline{2-3}
 &\multicolumn{2}{|c}{\multirow{3}{*}{${F^2} = \left( G_{\mu \upsilon }{y^\mu }{y^\upsilon } \right)^{\frac{1 + A_3}{1 + A_1}}\left( {{N_\mu }{y^\mu }} \right)^{ - 2\frac{A_3 + A_1}{1 + A_1}}$}} \\
 & \multicolumn{2}{|c}{}\\
 & \multicolumn{2}{|c}{}\\
 & \multicolumn{2}{|c}{no invariant metric in case of ${A_1} =  - 1$}\\
\hline

%\end{tabular}
%\end{center}
%\end{table}
%\end{widetext}

%\begin{widetext}
%\begin{table}
%\begin{center}
%\begin{tabular}{c|c|c}
%\hline
\multirow{13}{*}{$XDISIM2$} & \multirow{4}{*}{${N_\mu } $} & \multirow{2}{*}{${B_z}\left( \theta  \right):$}\\
& & \\
& &\multirow{2}{*}{${e^{\left( {1 + {A_3}} \right)\theta }}$} \\
& & \\
& & \\
\cline{2-3}
& \multirow{2}{*}{${H_{\left( {M,N} \right)\mu \upsilon }}$} & \multirow{2}{*}{${B_z}\left( \theta  \right):$} \\
 & & \\
 & \multirow{2}{*}{ $M =  - \frac{{1 + {A_3}}}{{1 + {A_1}}},N = \frac{{{A_1} - {A_3}}}{{1 + {A_1}}}$} & \multirow{2}{*}{${e^{2\left( {{A_3} - {A_1}} \right)\theta }}$}\\
 & & \\
\cline{2-3}
 &\multicolumn{2}{|c}{\multirow{3}{*}{${F^2} = {\left( {{H_{\left( {M,N} \right)\mu \upsilon }}{y^\mu }{y^\upsilon }} \right)^{\frac{{1 + {A_3}}}{{1 + {A_1}}}}}{\left( {{N_\mu }{y^\mu }} \right)^{ - 2\frac{{{A_3} - {A_1}}}{{1 + {A_1}}}}}$}} \\
 & \multicolumn{2}{|c}{}\\
 & \multicolumn{2}{|c}{}\\
 & \multicolumn{2}{|c}{a $t-z$ plane non-orthogonal linear}\\
 & \multicolumn{2}{|c}{transformation is made relative to $DISIM$}\\
\hline

\end{tabular}
\end{center}
\end{table}
%\end{widetext}

%\begin{widetext}
\begin{table}
\begin{center}
\begin{tabular}{c|c|c}
\hline
\multirow{6}{*}{$ISIM$} & \multirow{2}{*}{${N_\mu } $} & \multirow{2}{*}{${B_z}\left( \theta  \right):{e^{\theta }}$}\\
& & \\
\cline{2-3}
& \multirow{2}{*}{${G_{\mu \upsilon }}$} & \multirow{2}{*}{invariant} \\
 & & \\
\cline{2-3}
 &\multicolumn{2}{|c}{\multirow{2}{*}{${F^2} = {G_{\mu \upsilon }}{y^\mu }{y^\upsilon }$}} \\
 & \multicolumn{2}{|c}{}\\
\hline
$DIHOM$ & \multicolumn{2}{|c}{no invariant metric function}\\
\hline
$WDIHOM$ & \multicolumn{2}{|c}{the same as $DISIM$}\\
\hline
$IHOM$ & \multicolumn{2}{|c}{the same as $ISIM$}\\
\hline
$DTE1$ & \multicolumn{2}{|c}{no invariant metric function}\\
\hline
\multirow{10}{*}{$DTE2a1$} & \multirow{2}{*}{          ${N_\mu } $      } & \multirow{2}{*}{invariant}\\
& & \\
\cline{2-3}
& \multirow{4}{*}{${G_{\mu \upsilon }}$} & \multirow{2}{*}{${P_t}\left( \theta  \right), {P_z}\left( \theta  \right):$} \\
 & & \\
 & &\multirow{2}{*}{${e^{A_2\theta }}$} \\
 & & \\
\cline{2-3}
 &\multicolumn{2}{|c}{\multirow{2}{*}{$F = {N_\mu }{y^\mu }$}} \\
 & \multicolumn{2}{|c}{}\\
 & \multicolumn{2}{|c}{\multirow{2}{*}{the relation between $A_1$ and $A_2$: ${A_1} = {{A_2^2} \mathord{\left/
 {\vphantom {{A_2^2} 4}} \right.
 \kern-\nulldelimiterspace} 4}$}}\\
 & \multicolumn{2}{|c}{}\\
\hline
\multirow{12}{*}{$DTE2a2$} & \multirow{4}{*}{          ${N_\mu } $      } & \multirow{2}{*}{${P_t}\left( \theta  \right), {P_z}\left( \theta  \right):$}\\
& & \\
& & \multirow{2}{*}{${e^{\left( {2\lambda  - {A_2}} \right)\theta }}$}\\
& & \\
\cline{2-3}
& \multirow{4}{*}{${G_{\mu \upsilon }}$} & \multirow{2}{*}{${P_t}\left( \theta  \right), {P_z}\left( \theta  \right):$} \\
 & & \\
  & &  \multirow{2}{*}{${e^{2\lambda \theta }}$}\\
   & & \\
\cline{2-3}
 &\multicolumn{2}{|c}{\multirow{2}{*}{${F^2} = {\left( {{G_{\mu \upsilon }}{y^\mu }{y^\upsilon }} \right)^{\frac{{{A_2} - 2\lambda }}{{{A_2} - \lambda }}}}{\left( {{N_\mu }{y^\mu }} \right)^{\frac{{2\lambda }}{{{A_2} - \lambda }}}}$}} \\
 & \multicolumn{2}{|c}{}\\
 & \multicolumn{2}{|c}{the deform parameters satisfy: }\\
  & \multicolumn{2}{|c}{${\lambda ^2} - {A_2}\lambda  + {A_1} = 0$ and $\lambda  \ne {A_2}$}\\
\hline
$DTE2b$ & \multicolumn{2}{|c}{no invariant metric function}\\
\hline
$DTE3a$ & \multicolumn{2}{|c}{the same as $DTE2a$}\\
\hline
\multirow{11}{*}{$DTE3b$} & \multirow{2}{*}{          ${N_\mu } $      } & \multirow{2}{*}{invariant}\\
& & \\
\cline{2-3}
& \multirow{5}{*}{${H_{\left( {a,b} \right)\mu \upsilon }} = \left( {\begin{array}{*{20}{c}}
a&{}&{}&{a + b}\\
{}&b&{}&{}\\
{}&{}&b&{}\\
{a + b}&{}&{}&{a + 2b}
\end{array}} \right)$} & \multirow{5}{*}{invariant}\\
 & & \\
 & & \\
 & & \\
 & & \\
\cline{2-3}
 & \multicolumn{2}{|c}{\multirow{2}{*}{${F^2} = \prod\limits_{a,b} {{{\left( {{H_{\left( {a,b} \right)\mu \upsilon }}{y^\mu }{y^\upsilon }} \right)}^{{D_{a,b}}}}} $}}\\
 & \multicolumn{2}{|c}{}\\
 & \multicolumn{2}{|c}{\multirow{2}{*}{the constrain condition: $\sum\limits_{a,b} {{D_{a,b}}}  = 1$}}\\
 & \multicolumn{2}{|c}{}\\

\hline
%$TE(2)$ & \multicolumn{2}{|c}{the same as $DTE3b$ and hence $DTE3b$ denoted by $TE(2)$ %}\\
%\hline
\multirow{2}{*}{$TE(2)$} & \multicolumn{2}{|c}{the same as $DTE3b$ and hence }\\
 & \multicolumn{2}{|c}{$DTE3b$ denoted by $TE(2)$ }\\
\hline
$DISO(3)1$ & \multicolumn{2}{|c}{no invariant metric}\\
\hline
$DISO(3)2$ & \multicolumn{2}{|c}{no invariant metric}\\
\hline

\end{tabular}
\end{center}
\end{table}
%\end{widetext}

%\begin{widetext}
\begin{table}
\begin{center}
\begin{tabular}{c|c|c}
\hline
\multirow{11}{*}{$ISO(3)$} & \multirow{2}{*}{${T_\mu } = \left( {\begin{array}{*{20}{c}}
1&0&0&0
\end{array}} \right)^T$} & \multirow{2}{*}{invariant}\\
& & \\
\cline{2-3}
& \multirow{5}{*}{${G_{\left( {a,b} \right)\mu \upsilon }} = \left( {\begin{array}{*{20}{c}}
a&{}&{}&{}\\
{}&b&{}&{}\\
{}&{}&b&{}\\
{}&{}&{}&b
\end{array}} \right)$} & \multirow{5}{*}{invariant}\\
 & & \\
 & & \\
 & & \\
 & & \\
\cline{2-3}
 & \multicolumn{2}{|c}{\multirow{2}{*}{${F^2} = {\left( {{T_\mu }{y^\mu }} \right)^A}\prod\limits_{a,b} {{{\left( {{G_{\left( {a,b} \right)\mu \upsilon }}{y^\mu }{y^\upsilon }} \right)}^{{B_{a,b}}}}} $}}\\
 & \multicolumn{2}{|c}{}\\
 & \multicolumn{2}{|c}{\multirow{2}{*}{the constrain condition: $A + 2\sum\limits_{a,b} {{B_{a,b}}}  = 2$}}\\
 & \multicolumn{2}{|c}{}\\
 \hline
$DISO(2,1)1$ & \multicolumn{2}{|c}{no invariant metric}\\
\hline
$DISO(2,1)2$ & \multicolumn{2}{|c}{no invariant metric}\\
\hline
\multirow{11}{*}{$ISO(2,1)$} & \multirow{2}{*}{${X_\mu } = \left( {\begin{array}{*{20}{c}}
0&1&0&0
\end{array}} \right)^T$} & \multirow{2}{*}{invariant}\\
& & \\
\cline{2-3}
& \multirow{5}{*}{${\tilde G_{\left( {a,b} \right)\mu \upsilon }} = \left( {\begin{array}{*{20}{c}}
a&{}&{}&{}\\
{}&b&{}&{}\\
{}&{}&{ - a}&{}\\
{}&{}&{}&{ - a}
\end{array}} \right)$} & \multirow{5}{*}{invariant}\\
 & & \\
 & & \\
 & & \\
 & & \\
\cline{2-3}
 & \multicolumn{2}{|c}{\multirow{2}{*}{${F^2} = {\left( {{X_\mu }{y^\mu }} \right)^A}\prod\limits_{a,b} {{{\left( {{{\tilde G}_{\left( {a,b} \right)\mu \upsilon }}{y^\mu }{y^\upsilon }} \right)}^{{B_{a,b}}}}}$}}\\
 & \multicolumn{2}{|c}{}\\
 & \multicolumn{2}{|c}{\multirow{2}{*}{the constrain condition: $A + 2\sum\limits_{a,b} {{B_{a,b}}}  = 2$}}\\
 & \multicolumn{2}{|c}{}\\
 \hline
\hline

%\multirow{2}{*}{multi-rows} &
%\multicolumn{2}{|c|}{multi-columns} &
%\$multicolumn{2}{|c|}{\multirow{2}{*}{multi-row and col}} \\

%\cline{2-3}
%  \cline用于画横线 \cline{i-j}表示从第i列画到第j列
%& column-1 & column-2 & \multicolumn{2}{|c}{} \\
%\hline

\end{tabular}
\end{center}
\end{table}
%\end{widetext}

Note that the metric function which is invariant under the deformed Poincar\'e subgroup is always in the form
\beq \label{eq:metricgeneral}
{F^2} = {\left( {{A_\mu }{y^\mu }} \right)^{2 - 2\sum\limits_{a,b} {{D_{a,b}}} }}\prod\limits_{a,b} {{{\left( {{B_{\left( {a,b} \right)\mu \upsilon }}{y^\mu }{y^\upsilon }} \right)}^{{D_{a,b}}}}} ,
\eeq
where $A_\mu$ can be one of  $N_\mu$,  $T_\mu$ or  $X_\mu$ while ${B_{\left( {a,b} \right)\mu \upsilon }}$ can be one of ${G_{\left( {a,b} \right)\mu \upsilon }}$, ${\tilde G_{\left( {a,b} \right)\mu \upsilon }}$ or ${H_{\left( {a,b} \right)\mu \upsilon }}$ and the combination of $A_\mu$ and ${B_{\left( {a,b} \right)\mu \upsilon }}$ is different for different group. The forms frequently  appear are
\beq \label{eq:metricformfrequently}
\begin{array}{c}
  {F^2} = {G_{\mu \upsilon }}{y^\mu }{y^\upsilon } \\
  {\left( {{N_\mu }{y^\mu }} \right)^2} \\
  {\left( {{G_{\mu \upsilon }}{y^\mu }{y^\upsilon }} \right)^{1 - A}}{\left( {{N_\mu }{y^\mu }} \right)^{2A}}.
\end{array}
\eeq

Among the undeformed groups, only the $ISIM$ group invariant metric function is the Minkowski one while the $TE(2)$, $ISO(3)$ and $ISO(2)$ invariant metric functions are all form of \eqref{eq:metricgeneral}, where $A_\mu$ is $N_\mu$,  $T_\mu$ and  $X_\mu$ and ${B_{\left( {a,b} \right)\mu \upsilon }}$ is ${H_{\left( {a,b} \right)\mu \upsilon }}$, ${G_{\left( {a,b} \right)\mu \upsilon }}$ and ${\tilde G_{\left( {a,b} \right)\mu \upsilon }}$ respectively.

\subsection{More forms of the metric functions}

Invariant metric function form like \eqref{eq:metricgeneral} is representative. However, the invariant metric function can have more plenty of forms available, e.g. the invariant metric function of $DISIM$ is
\beq \label{eq:mtrorgidisim}
{F^2} = {\left( {{G_{\mu \upsilon }}{y^\mu }{y^\upsilon }} \right)^{1 + {A_2}}}{\left( {{N_\mu }{y^\mu }} \right)^{ - 2{A_2}}}
\eeq
in general, it is also allowable to have the metric function of the form
\beq \label{eq:absmtridisim}
{F^2} = {\left| {{G_{\mu \upsilon }}{y^\mu }{y^\upsilon }} \right|^{1 + {A_2}}}{\left| {{N_\mu }{y^\mu }} \right|^{ - 2{A_2}}}
\eeq
 even that of the form
\beq \label{eq:sgnmtridisim}
{F^2} = {\mathop{\rm sgn}} \left( {{G_{\mu \upsilon }}{y^\mu }{y^\upsilon }} \right){\left| {{G_{\mu \upsilon }}{y^\mu }{y^\upsilon }} \right|^{1 + {A_2}}}{\left| {{N_\mu }{y^\mu }} \right|^{ - 2{A_2}}}.
\eeq
However, compare to \eqref{eq:mtrorgidisim} and  \eqref{eq:absmtridisim}, \eqref{eq:sgnmtridisim} is better for there may be some region where ${G_{\mu \upsilon }}{y^\mu }{y^\upsilon } < 0$ or ${N_\mu }{y^\mu } < 0$, \eqref{eq:mtrorgidisim} is not well defined when the deformation parameter $A_2$ is not an integer while \eqref{eq:absmtridisim} distinguishes light-like, time-like and space-like totally. \eqref{eq:sgnmtridisim} is therefor well defined in pseudo-Finsler spacetime.

Moreover, metric function may have plenty of additional structure. If there exist some scalar function $\phi \left( {{y^\mu }} \right)$, which is invariant under the group action and is the zero degree homogenous function of ${y^\mu }$, then the product of $\phi$ and the metric function is still an invariant metric function, e.g.
\beq \label{eq:expofphi}
{F^2} = {G_{\mu \upsilon }}{y^\mu }{y^\upsilon }\cos \left( {\omega \frac{{{T_\mu }{y^\mu }}}{{{X_\mu }{y^\mu }}} + \theta } \right),
\eeq
where $\omega$ and $\theta$ are arbitrary parameters.
This kind of metric function is allowable in spacetime only possessing rotational symmetry in $y-z$ plane.

To find out the most general form of invariant metric function, one need to determine all such kind of invariant zero degree homogenous function $\phi$ of ${y^\mu }$. Table II lists the invariant zero degree homogenous function $\phi$ of ${y^\mu }$ for various of Poincar\'e subgroups and their deformed partners.

\begin{table}
\begin{center}
\caption{invariant zero degree functions of Poincar\'e subgroup and the deformed partner}
\begin{tabular}{c|c}
\hline
\hline
symmetric & \multirow{2}{*}{invariant zero degree homogenous function $\phi$}\\
group & \\
\hline
\hline
DISIM & $\phi$=1\\
\hline
XDISIM1 & $\phi$=1\\
\hline
XDISIM2 & $\phi$=1\\
\hline
ISIM & $\phi$=1\\
\hline
DTE2a1 & $\phi$=1\\
\hline
DTE2a2 & $\phi$=1\\
\hline
\multirow{2}{*}{DTE3b} & \multirow{2}{*}{$\phi  = \frac{{{{\left( {{N_\mu }{y^\mu }} \right)}^2}}}{{{G_{\mu \upsilon }}{y^\mu }{y^\upsilon }}}$}\\
 & \\
\hline
\multirow{2}{*}{TE(2)} & \multirow{2}{*}{${\phi _{a,b;c,d}} = \frac{{{H_{\left( {a,b} \right)\mu \upsilon }}{y^\mu }{y^\upsilon }}}{{{H_{\left( {c,d} \right)\mu \upsilon }}{y^\mu }{y^\upsilon }}}$}\\
 & \\
 \hline
\multirow{2}{*}{ISO(3)} & \multirow{2}{*}{${\phi _{a,b}} = \frac{{{{\left( {{T_\mu }{y^\mu }} \right)}^2}}}{{{G_{\left( {a,b} \right)\mu \upsilon }}{y^\mu }{y^\upsilon }}}$ and ${\phi _{a,b;c,d}} = \frac{{{G_{\left( {a,b} \right)\mu \upsilon }}{y^\mu }{y^\upsilon }}}{{{G_{\left( {c,d} \right)\mu \upsilon }}{y^\mu }{y^\upsilon }}}$}\\
 & \\
 \hline
\multirow{2}{*}{ISO(2,1)} & \multirow{2}{*}{${\phi _{a,b}} = \frac{{{{\left( {{X_\mu }{y^\mu }} \right)}^2}}}{{{{\tilde G}_{\left( {a,b} \right)\mu \upsilon }}{y^\mu }{y^\upsilon }}}$ and ${\phi _{a,b;c,d}} = \frac{{{{\tilde G}_{\left( {a,b} \right)\mu \upsilon }}{y^\mu }{y^\upsilon }}}{{{{\tilde G}_{\left( {c,d} \right)\mu \upsilon }}{y^\mu }{y^\upsilon }}}$}\\
 & \\
 \hline
 \hline
\end{tabular}
\end{center}
\end{table}

For DTE3b, TE(2), ISO(3), ISO(2,1), the invariant metric function can take the form of
\beq \label{eq:addmetricform}
\left\{ {\begin{array}{*{20}{c}}
{\rm{DTE3b}}:&{F^2} = {{\left( {{G_{\mu \upsilon }}{y^\mu }{y^\upsilon }} \right)}^{1 - A}}{{\left( {{N_\mu }{y^\mu }} \right)}^{2A}}\\
 & \cdot S\left( {{\phi _{{\rm{DTE3b}}}}} \right)\\
{{\rm{TE}}\left( {\rm{2}} \right):}&{F^2} = \left[ {\prod\limits_{a,b} {{{\left( {{H_{\left( {a,b} \right)\mu \upsilon }}{y^\mu }{y^\upsilon }} \right)}^{{D_{a,b}}}}} } \right]\\
 &\cdot  S\left( {{\phi _{{\rm{TE}}\left( 2 \right)c,d;e,f}}} \right)\\
{\begin{array}{*{20}{c}}
{{\rm{ISO}}\left( {\rm{3}} \right):}\\
{}
\end{array}}&{\begin{array}{*{20}{c}}
{{F^2} =\left[ {\prod\limits_{a,b} {{{\left( {{G_{\left( {a,b} \right)\mu \upsilon }}{y^\mu }{y^\upsilon }} \right)}^{{B_{a,b}}}}} } \right]}\\
\cdot {{{\left( {{T_\mu }{y^\mu }} \right)}^A}S\left( {{\phi _{{\rm{ISO}}\left( {\rm{3}} \right)c,d}},{\phi _{{\rm{ISO}}\left( {\rm{3}} \right)c,d;e,f}}} \right)}
\end{array}}\\
{\begin{array}{*{20}{c}}
{{\rm{ISO}}\left( {{\rm{2,1}}} \right):}\\
{}
\end{array}}&{\begin{array}{*{20}{c}}
{{F^2} =\left[ {\prod\limits_{a,b} {{{\left( {{{\tilde G}_{\left( {a,b} \right)\mu \upsilon }}{y^\mu }{y^\upsilon }} \right)}^{{B_{a,b}}}}} } \right]}\\
\cdot {{{\left( {{X_\mu }{y^\mu }} \right)}^A}S\left( {{\phi _{{\rm{ISO}}\left( {{\rm{2,1}}} \right)c,d}},{\phi _{{\rm{ISO}}\left( {{\rm{2,1}}} \right)c,d;e,f}}} \right)},
\end{array}}
\end{array}} \right.
\eeq
where $S$ is an arbitrary function.

We thus give all of the Finsler-Minkowski type metric functions which corresponding the 3 generators and four generators Lorentz subgroups and their deformed partner.

Note that the structure of given metric functions is the product of several parts as the ansatz of \eqref{eq:finmtr} to enable the group action invariance and the degree 2 homogeneity. Actually, if there are several such groups of components, we can construct metric function by adding different parts constructing from different groups of components together,

\beq \label{eq:addmetricformtoge}
F = \sum\limits_n {{F_{\left( n \right)}}},
\eeq
where $F_{\left( n \right)}^n = \sum\limits_i {F_i^n}$. The Poincar\'e subgroups in \eqref{eq:addmetricform} are just those which have more than one group of components to construct the invariant metric function. Therefore they can possess the metric function of the form in \eqref{eq:addmetricformtoge}. For example, $TE(2)$ can have such kind of metric function as
\beq \label{eq:mtrofadding}
F = A\sqrt {{G_{\mu \upsilon }}{y^\mu }{y^\upsilon } + {{\left( {{N_\mu }{y^\mu }} \right)}^2}}  + B\sqrt {{G_{\mu \upsilon }}{y^\mu }{y^\upsilon }}  + C{N_\mu }{y^\mu }.
\eeq

\subsection{Conclusion and Outlook}

We have obtained all possible group action invariant Finsler-Minkowski metric functions for the semi-product group of three generators and four generators Lorentz subgroups with translation group $T(4)$ and their deformed partner. We find that the group action invariance has strong restrictions on the possible metric functions such that the invariant metric functions for different groups may have the same form. Finally there are only several kinds of metric functions corresponding to a few of symmetry groups which is listed in the Table III, where we only list the maximal symmetric group if there are several groups that correspond to the same metric functions.
\begin{table}
\begin{center}
\caption{Various possible symmetric groups and their corresponding invariant metric functions}
\begin{tabular}{c|c}
\hline
\hline
symmetry  & \multirow{2}{*}{metric function}\\
group & \\
\hline
\hline
\multirow{3}{*}{DISIM} & \multirow{2}{*}{${F^2} = {\mathop{\rm sgn}} \left( {{G_{\mu \upsilon }}{y^\mu }{y^\upsilon }} \right){\left| {{G_{\mu \upsilon }}{y^\mu }{y^\upsilon }} \right|^{1 - A}}{\left| {{N_\mu }{y^\mu }} \right|^{2A}}$}\\
 & \\
 & $A$ is given by individual deformation\\
\hline
Poincar\'e & \multirow{2}{*}{${F^2} = {G_{\mu \upsilon }}{y^\mu }{y^\upsilon }$}\\
(special case & \\
of DISIM)& the case of $A=0$\\
\hline
DTE2a1 & \multirow{2}{*}{$F = \left| {{N_\mu }{y^\mu }} \right|$}\\
(special case & \\
of DISIM)& the case of $A=1$\\
\hline
 & \multirow{2}{*}{$F = \sum\limits_N {{D_N}{F_{\left( N \right)}}} $ where $F_{\left( N \right)}^N = \sum\limits_i {{C_i}F_i^N} $}\\
  & \\
TE(2) & ${F_i} \in \left\{ {f\left| {f = \sqrt {\prod\limits_{a,b} {{{\left( {{H_{\left( {a,b} \right)\mu \upsilon }}{y^\mu }{y^\upsilon }} \right)}^{{D_{a,b}}}}} } } \right.} \right.$\\
 (DTE3b)& $\left. \cdot{S\left( {{\phi _{{\rm{TE}}\left( 2 \right)c,d;e,f}}} \right)} \right\}$\\
 & \multirow{2}{*}{$\sum {{D_{a,b}}}  = 1$}\\
  &  \\
\hline
\multirow{6}{*}{ISO(3)}& \multirow{2}{*}{$F = \sum\limits_N {{D_N}{F_{\left( N \right)}}} $ where $F_{\left( N \right)}^N = \sum\limits_i {{C_i}F_i^N} $}\\
  & \\
  & ${F_i} \in \left\{ {f\left| {f = {{\left( {{T_\mu }{y^\mu }} \right)}^A}\left[ {\prod\limits_{a,b} {{{\left( {{G_{\left( {a,b} \right)\mu \upsilon }}{y^\mu }{y^\upsilon }} \right)}^{{B_{a,b}}}}} } \right]} \right.} \right.$\\
  & $\cdot\left. {S\left( {{\phi _{{\rm{ISO}}\left( {\rm{3}} \right)c,d}},{\phi _{{\rm{ISO}}\left( {\rm{3}} \right)c,d;e,f}}} \right)} \right\}$\\
 & \multirow{2}{*}{$A + 2\sum {{B_{a,b}}}  = 2$}\\
  &  \\
\hline
\multirow{6}{*}{ISO(2,1)}& \multirow{2}{*}{$F = \sum\limits_N {{D_N}{F_{\left( N \right)}}} $ where $F_{\left( N \right)}^N = \sum\limits_i {{C_i}F_i^N} $}\\
  & \\
  & ${F_i} \in \left\{ {f\left| {f = {{\left( {{X_\mu }{y^\mu }} \right)}^A}\left[ {\prod\limits_{a,b} {{{\left( {{{\tilde G}_{\left( {a,b} \right)\mu \upsilon }}{y^\mu }{y^\upsilon }} \right)}^{{B_{a,b}}}}} } \right]} \right.} \right.$\\
  & $\cdot\left. {S\left( {{\phi _{{\rm{ISO}}\left( {{\rm{2}},{\rm{1}}} \right)c,d}},{\phi _{{\rm{ISO}}\left( {{\rm{2}},{\rm{1}}} \right)c,d;e,f}}} \right)} \right\}$\\
 & \multirow{2}{*}{$A + 2\sum {{B_{a,b}}}  = 2$}\\
  &  \\
\hline
\hline
\end{tabular}
\end{center}
\end{table}

It can be observed that the undeformed semi-product group usually has richer Finsler structures than the deformed one. The invariant metric corresponding to deformed group uaually has larger symmetry. The Finsler metric which corresponds to the largest Poincar\'e group is unique while the Finsler structure corresponding to the next to largest symmetry group DISIM is determined uniquely by the deformation parameter of DISIM itself. The semi-product group of three generators Lorentz subgroups have much richer Finsler structure, which can not be determined uniquely even constrained in Riemann geometry. We argue that a reasonable rotation operation should not have the additional accompanied scale transformation, i.e. the the Lorentz violation should not happen in the rotation sector but in the boost sector, in our previous paper \cite{Zhang:2012ty}. The investigation on the invariant metric function indicates that the existence of invariant metric function automatically excludes the additional accompanied scale transformation for rotation operation, i.e. it is a requirement of geometry that the rotation operation is kept even in a Lorentz violation theory.

In our next subsequent paper, we investigate the single particle dynamics and the field theories in the obtained various kind of Finsler-Minkowski spacetime. It reveals that there is the fractional(even irrational) power of derivatives problem both in the single particle dynamics and in the dynamics of field theories. In field theory, we expand the theory according to the power in deformation parameters and result in the lagrangian an non-local log term.

Though the geometries we obtain in this paper have a large freedom, the dynamics in the corresponding spacetime seems supplying constrains on the possible form of geometries. We leave the problem in our next paper.

\bigskip

\acknowledgments{}
X.Xue wish to thank Hanqing Zheng for illuminating discussions.

%\appendix
%\section{Geodesic equation}
%\label{ap:geodesic}

\bibliographystyle{unsrt}

\end{document}